# Emergent quantum phenomena via phase-coherence engineering in infinite-layer nickelate superconductors

Haoran Ji[1#], Zheyuan Xie[1#], Xiaofang Fu[2#], Zihan Cui[3,4], Minghui Xu[2], Guang-Ming Zhang[5,6], Yi-feng Yang[7,8,9], Haiwen Liu[10,11,12], Yi Liu[3,4*], Liang Qiao[2*] & Jian Wang[1,13,14*]

[1]International Center for Quantum Materials, School of Physics, Peking University, Beijing 100871, China

[2]School of Physics, University of Electronic Science and Technology of China, Chengdu 610054, China

[3]School of Physics and Beijing Key Laboratory of Opto-electronic Functional Materials & Micro-nano Devices, Renmin University of China, Beijing 100872, China

[4]Key Laboratory of Quantum State Construction and Manipulation (Ministry of Education), Renmin University of China, Beijing 100872, China

[5]State Key Laboratory of Quantum Functional Materials, School of Physical Science and Technology, ShanghaiTech University, Shanghai 201210 China

[6]Department of Physics, Tsinghua University, Beijing 100084, China

[7]Beijing National Laboratory for Condensed Matter Physics and Institute of Physics Chinese Academy of Sciences, Beijing 100190, China

[8]University of Chinese Academy of Sciences, Beijing 100049, China

[9]Songshan Lake Materials Laboratory, Dongguan, Guangdong 523808, China

[10]Center for Advanced Quantum Studies, Department of Physics, Beijing Normal University, Beijing 100875, China

[11]Key Laboratory of Multiscale Spin Physics, Ministry of Education, Beijing 100875, China

[12]Interdisciplinary Center for Theoretical Physics and Information Sciences, Fudan University, Shanghai 200433, China

[13]Collaborative Innovation Center of Quantum Matter, Beijing 100871, China

[14]Hefei National Laboratory, Hefei 230088, China

[#]These authors contribute equally.
Correspondence to: jianwangphysics@pku.edu.cn (J.W.),
liang.qiao@uestc.edu.cn (L.Q.),
yiliu@ruc.edu.cn (Y.L.),

**Dimensionality of a physical system, conventionally an invariant geometric characteristic, fundamentally governs the universality class of phase transitions and the landscape of emergent collective phenomena. In low-dimensional or layered high-temperature superconductors, the macroscopic phase coherence of superconducting orders is typically confined in two dimensions, underscoring the critical role of phase fluctuations in determining the overall phase diagrams. Here, we strategically enhance the phase fluctuations by fabricating periodically arranged nano-holes in the infinite-layer nickelate superconducting films, effectively constructing Josephson junction arrays. In the nano-patterned films, the weakening of macroscopic phase coherence drives a two-stage superconducting transition towards an anomalous metallic ground state with saturated resistance. The emergence of charge-2e quantum oscillations manifests the coherence across the array, while an anomalous zero-field magnetoresistance peak signifies the extreme quantum phase fluctuations persisting to ultralow temperatures. Remarkably, with quantum fluctuations enhanced synergistically by nano-patterning and magnetic fields, an anomalous reversal of superconducting anisotropy is observed in Nd-nickelates, where in-plane critical fields fall below out-of-plane values ($B_{c/\!/} < B_{c\perp}$). The evolution of anisotropy may unmask an internal exchange-Zeeman field coupled to the collective electronic states. Our results unveil how superconductivity evolves in response to phase fluctuations, establishing nano-patterning as a powerful paradigm to uncover hidden intertwined orders in strongly correlated systems.**

Superconductivity arises from two fundamental processes: the formation of Cooper pairs and the establishment of phase coherence among them. In conventional three-dimensional (3D) superconductors described by the Bardeen-Cooper-Schrieffer (BCS) theory, the long-range phase coherence is established simultaneously with Cooper pairing, as the phase-stiffness is comparably strong. However, in superconducting systems with reduced dimensionality or low carrier density, the phase stiffness is significantly reduced, and the superconducting transition is essentially dominated by phase fluctuations[1,2]. For instance, in two-dimensional (2D) systems, the phase coherent superconducting state is characterized by the Berezinskii-Kosterlitz-Thouless (BKT) transition[3,4], which corresponds to the binding of vortex and antivortex, the phase singularities of superconducting order parameter. Originating from the quantum phase fluctuations approaching zero-temperature limit, the anomalous metallic state emerges as an unexpected ground state in 2D superconducting system[5-9]. The emergence of the anomalous metallic state violates the expectation derived from the Heisenberg uncertainty principle in 2D systems for Cooper pairs, where the ground states should be either a superconducting state with macroscopic phase coherence or an insulating state without coherence[10].

In layered unconventional high-transition-temperature (high-$T_c$) superconducting systems, such as cuprate superconductors, the superconducting phase fluctuations play a crucial role in governing the unconventional superconductivity as well as the global phase diagram with diverse strongly correlated phases[2]. The relatively small phase stiffness, stemming from the low carrier density, determines the superconducting phase boundary in the underdoped region[2,11]. The weakly coupled $CuO_2$ planes and the Cu-$d_{x^2-y^2}$ derived Fermi surface account for the two-dimensionality of the cuprates, suggesting the importance of reduced dimensionality in high-$T_c$ superconductivity[12].

As structural and electronic analogues to the cuprate superconductor, the recently discovered nickelate family of unconventional high-$T_c$ superconductors[13-19], especially the infinite-layer nickelate $R_{1-x}Sr_xNiO_2$ ($R$ = La, Pr, Nd) thin films[18,20-23], offer a promising platform to unravel the shared fundamental mechanisms of unconventional high-$T_c$ superconductivity. This naturally leads to the question: what is the role of phase fluctuations in the nickelate superconductors? The infinite-layer structure corresponds to $n = \infty$ member of the layered family[24] $R_{n+1}(NiO_2)_nO_2$, where the blocking layers mediating the coupling between adjacent units are minimized to active $R^{3+}$ cations (Fig. 1**a**). Notably, the Fermi surface of $R_{1-x}Sr_xNiO_2$ films consists of a quasi-2D Ni-$d_{x^2-y^2}$ hole pocket and a 3D electron pocket[25,26]. The composite dimensionality of crystal and electronic structures blurs the superconducting dimensionality of the $R_{1-x}Sr_xNiO_2$ films, triggering intensive debates ranging from isotropic[27,28], quasi-2D anisotropic[29] critical fields, and even 2D to 3D crossover[30-32]. Therefore, the nickelate superconductors should provide a distinctive framework, which serves to elucidate the intricate interplay between unconventional superconductivity and various forms of emergent collective fluctuations.

With these motivations, we fabricate the Josephson junction array (JJA)-patterned infinite-layer nickelate superconductor $Nd_{0.8}Sr_{0.2}NiO_2$ thin films and investigate the electrical transport properties. By introducing periodically arranged nano-holes to artificially construct the JJA, we provide a systematic characterization on the evolution of unconventional superconductivity under progressively enhanced phase fluctuations.

How the superconductivity evolves and manifests in response to phase fluctuation modulations in the nano-patterned $Nd_{0.8}Sr_{0.2}NiO_2$ films provides critical insights into the role of phase coherence in stabilizing the global superconducting states, which would advance the understanding and application of the unconventional superconductivity.

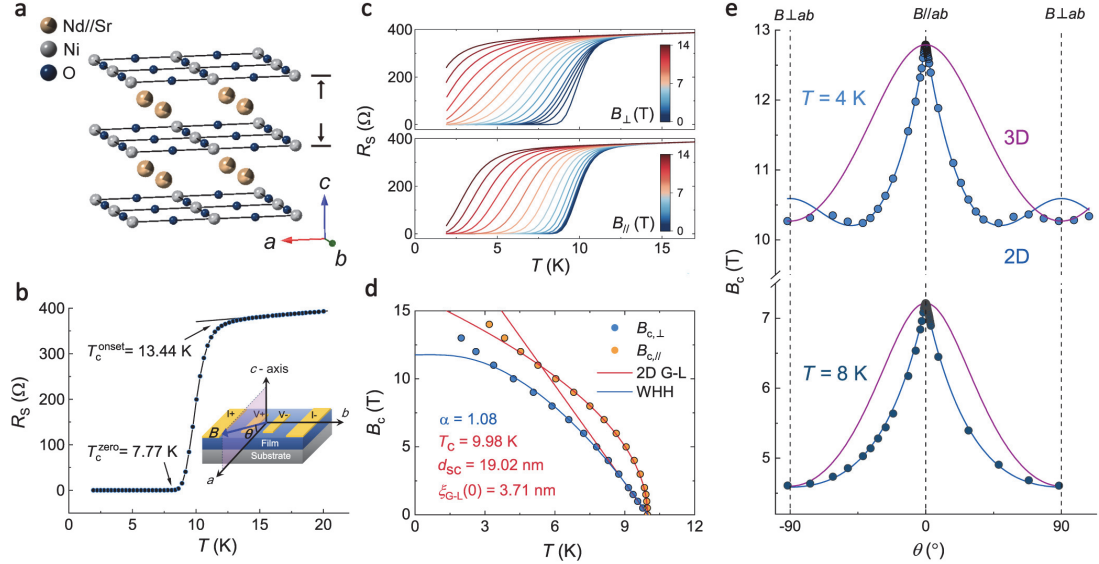

**Fig. 1 | Dimensionality of superconductivity in unpatterned $Nd_{0.8}Sr_{0.2}NiO_2$ film. a**, Crystal structure of the infinite-layer nickelate $Nd_{0.8}Sr_{0.2}NiO_2$. **b**, Temperature-dependent sheet resistance $R_S(T)$ at zero magnetic field from 2 K to 20 K. Inset: schematic of standard four-electrode transport measurement configuration. **c**, $R_S(T)$ curves under out-of-plane ($B \perp ab$ plane, denoted as $\perp$ in the upper panel) and in-plane ($B \parallel ab$ plane, denoted as $\parallel$ in the lower panel) magnetic fields. **d**, Temperature-dependent upper critical fields $B_c(T)$ for out-of-plane and in-plane directions. Red solid lines are the 2D G-L fits of the $B_c(T)$ near $T_c$. Blue solid line represents the WHH fit. **e**, Polar angular dependent upper critical fields $B_c(\theta)$ at $T = 4$ K (upper panel) and $T = 8$ K (lower panel). Blue and purple solid lines represent the fits with 2D Tinkham model and 3D G-L model, respectively. Vertical dashed lines mark the in-plane ($B \parallel ab$) and out-of-plane ($B \perp ab$) directions.

To start with, the standard four-electrode method is used to characterize the electric transport properties of a 15-nm-thick unpatterned $Nd_{0.8}Sr_{0.2}NiO_2$ thin film. Figure 1**b** displays the temperature-dependent sheet resistance $R_S(T)$, which shows a superconducting transition beginning at $T_c^{onset} = 13.44$ K, and a zero-resistance state below $T_c^{zero} = 7.77$ K. To evaluate the superconducting dimensionality of Nd$_{0.8}$Sr$_{0.2}$NiO$_2$ films, the resistive characteristics are investigated with magnetic fields applied along different polar angle $\theta$, where $\theta$ corresponds to the angle between the magnetic field and the *ab*-plane of Nd$_{0.8}$Sr$_{0.2}$NiO$_2$ film (inset of Fig. 1**b**). Figure 1**c** displays the $R_S(T)$ curves measured under different out-of-plane ($\theta = 90°$) and in-plane ($\theta = 0°$) magnetic fields, showing discernible signatures of anisotropy. The temperature-dependent in-plane upper critical fields $B_{c\parallel}(T)$ and out-of-plane upper critical fields $B_{c\perp}(T)$ are shown in Fig. 1**d**, which are defined as the magnetic fields required to reach 50% of the normal state sheet resistance $R_N$. Near $T_c$, the temperature-dependences of both $B_{c\perp}$ and $B_{c\parallel}$ are well described by the phenomenological Ginzburg-Landau (G-L) formula[33] for 2D superconductors based on the orbital pair-breaking effect under 2D confinement (Fig. 1**d**). The Werthamer-Helfand-Hohenberg (WHH) theory[34] can fit $B_{c\perp}(T)$ over a broad temperature range, which yields a Maki parameter $\alpha \equiv \sqrt{2}B_{\text{orbital}} / B_{\text{Pauli}} = 1.08$ (assuming negligible spin-orbital interaction, with $B_{\text{orbital}}$ and $B_{\text{Pauli}}$ denoting the orbital limit and Pauli paramagnetic limit, respectively), and thereby implies the comparable contribution from paramagnetic effect and orbital effect (Fig. 1**d**).

Figure 1**e** represents the polar angular dependent upper critical fields $B_c(\theta)$. At $T = 8$ K, a relatively high temperature near $T_c$, $B_c(\theta)$ increases monotonically from $|\theta| = 90°$ ($B \perp ab$) to $\theta = 0°$ ($B \parallel ab$), as shown in the bottom of Fig. 1**e**. Among the entire $\theta$ range, $B_c(\theta)$ manifests a cusp-like peak centered at $\theta = 0°$, which could be described by the 2D Tinkham model[35] but deviates from the 3D G-L model. This fit ascribes the anisotropy of $B_c(\theta)$ to the orbital effect with dominant 2D nature, consistent with the thin-film geometry. However, when the temperature decreases to 4 K, $B_c(\theta)$ exhibits slight non-monotonicity from $|\theta| = 90°$ ($B \perp ab$) to $\theta = 0°$ ($B \parallel ab$) with minima at $|\theta| \sim 42°$ (top in Fig. 1**e**). While the 2D Tinkham model captures the overall anisotropic $B_c(\theta)$, the anisotropy ratio $\gamma = B_{c\parallel}/B_{c\perp}$ is merely ~124% at 4 K (~157% at 8 K), substantially smaller than those in conventional 2D superconductors. For comparison, $\gamma$ value of La$_{0.8}$Sr$_{0.2}$NiO$_2$ thin film is ~524% at 4 K (Extended Data Fig. 4). Therefore, from the perspective of the critical field anisotropy, the La$_{0.8}$Sr$_{0.2}$NiO$_2$ films can be considered as quasi-2D superconductors aligned with their geometric two-dimensionality, whereas the Nd$_{0.8}$Sr$_{0.2}$NiO$_2$ films are characterized by a complex superconducting dimensionality. Intuitively, the predominant 2D nature of La-based films arises from the larger interlayer spacing between the adjacent NiO$_2$ layers separated by La$^{3+}$ (ionic radius $r^{\text{La}^{3+}} \sim 117.2$ pm, *c*-axis lattice constant ~ 3.44 Å) than Nd$^{3+}$ ($r^{\text{Nd}^{3+}} \sim 112.3$ pm, *c*

~ 3.36 Å) [31,36-38]. Differing from both cuprates and other nickelate superconductors, the $Nd_{0.8}Sr_{0.2}NiO_2$ films may imply the involvement of both orbital effect and possible spin-configuration-dependent paramagnetic effect, which have different dimensional characteristics but comparable energy scales.

In order to enhance the phase fluctuations and investigate the associated phase transitions in a controllable manner, we employed the nano-pattern engineering of Josephson-junction array (JJA). Generally, JJA could be constructed by patterning superconducting films with periodic holes[39,40] or embedding superconducting islands on normal metal[41-43]. By tuning parameters such as the etching time[40], geometric scales (e.g., size and spacing of the superconducting islands)[41], or gate voltage[42-44], one can effectively modulate the phase ordering of the global superconducting state. Here, we nano-patterned the $Nd_{0.8}Sr_{0.2}NiO_2$ thin films with periodic holes array by reactive ion etching (RIE) through a transferred anodized aluminum oxide (AAO) membrane mask (Fig. 2**a**). The holes, ~70 nm in diameter ($d$), are arranged in a triangular distribution with a center-to-center spacing ($D$) of ~125 nm (Fig. 3**a**). In this configuration, the un-etched regions surrounded by three nanoholes act as superconducting islands, and their connections serve as the Josephson links, thereby forming an effective JJA system[40].

Figure 2**b** shows a typical $R_S(T)$ curve of the nano-patterned $Nd_{0.8}Sr_{0.2}NiO_2$ film with 10 seconds (10s)-etching time, which exhibits a two-stage superconducting transition spanning over a wide temperature regime. The two-stage transitions have been widely studied as a characteristic phenomenon in 2D superconducting JJA system[41], where the first stage is ascribed to the establishment of local superconductivity within individual islands (around $T_1$), and the second stage to the inter-island phase coherence across the links (around $T_2$, also as schematically illustrated in Fig. 3**a**). The characteristic temperatures for the local superconductivity $T_1$, and inter-island phase coherence $T_2$ are identified by the maxima in the second derivative ($-d^2R_S/dT^2$) of the $R_S(T)$ curves[44]. As the etching time increases from 5s to 15s, the $R_S(T)$ curves exhibit a tendency where the inter-island phase coherence is progressively suppressed toward lower temperatures, showcasing the nano-patterning-dependent enhancement of phase fluctuations (Supplementary Information Fig. S1).

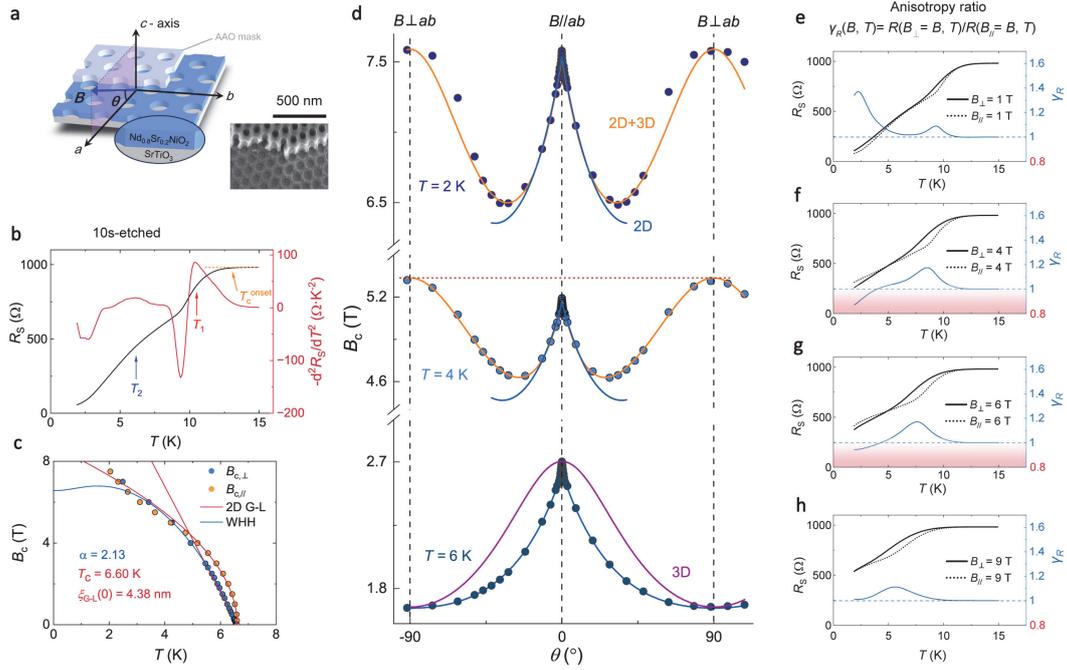

**Fig. 2 | Reversal of superconducting anisotropy in 10s-etched nano-patterned Nd$_{0.8}$Sr$_{0.2}$NiO$_2$ film. a**, Schematic (left) and scanning electron microscope (SEM) image (right) of the nano-patterned nickelate film on a substrate with a residual piece of AAO mask on top. **b**, $R_S(T)$ curve and second derivative ($-d^2R_S/dT^2$) curve of the two-stage transition, marking the onset transition at $T_c^{onset}$, establishment of local superconductivity around $T_1$, and inter-island phase coherence around $T_2$. **c**, $B_{c\perp}(T)$, $B_{c\parallel}(T)$, and the corresponding 2D G-L fits (red solid lines) and WHH fit (blue solid line). **d**, $B_c(\theta)$ at $T = 2$ K (top), 4 K (middle) and $T = 6$ K (bottom). Purple and blue solid lines represent the fits with 3D G-L model and 2D Tinkham model, respectively. Orange solid lines correspond to the fits using a combination of 2D and 3D models. Red dotted line highlights the superconducting anisotropy reversal where $B_{c\parallel} < B_{c\perp}$. **e-h**, $R_S(T)$ curves under different in-plane (dotted line) and out-of-plane (solid line) magnetic fields of 1 T (**e**), 4 T (**f**), 6 T (**g**), and 9 T (**h**). The blue curves represent the temperature-dependent resistance anisotropy ratio $\gamma_R$, which correspond to the ratio between the two $R_S(T)$ curves under orthogonal magnetic fields, $\gamma_R(|B|, T) = R_S(B_\perp = |B|, T)/R_S(B_\parallel = |B|, T)$. Blue dashed line marks $\gamma_R = 1$, below which the superconductivity is more fragile to in-plane magnetic fields ($B_{c\parallel} < B_{c\perp}$, highlighted by the red background).

Then, the evolution of superconducting anisotropy upon the enhanced fluctuations is systematically investigated. In the 10s-etched nano-patterned Nd$_{0.8}$Sr$_{0.2}$NiO$_2$ film, the $B_{c\perp}(T)$ and $B_{c\parallel}(T)$ curves converge, and even cross below ~ 4 K (Fig. 2**c**), signifying the unconventional superconducting anisotropy. The $B_{c\perp}(T)$ and $B_{c\parallel}(T)$ curves near $T_c$ could be fitted by the 2D G-L formula (Fig. 2**c**), and the $B_c(\theta)$ curves at 6 K are captured by 2D Tinkham model (bottom of Fig. 2**d**), collectively indicating the orbital effect with dominant 2D character close to $T_c$. The Maki parameter $\alpha$ increases monotonically with increasing etching time, tilting the competition between the orbital and paramagnetic pair-breaking effects, with the latter one becoming increasingly dominant (Fig. 2**c**, $\alpha$ with different etching times can be found in Extended Data Fig. 1). Remarkably, $B_c(\theta)$ curves at 2 K and 4 K exhibit dramatic non-monotonicity from $|\theta| = 90°$ ($B\perp$ab) to $\theta = 0°$ ($B\parallel$ab) (top and middle of Fig. 2**d**). This non-monotonicity manifests a striking reversal of superconducting anisotropy that is characterized by $B_{c\parallel} < B_{c\perp}$, completely opposite to $B_{c\perp} < B_{c\parallel}$ typically expected for superconductivity under 2D geometric confinement. While the conventional 2D model reproduces the $B_c(\theta)$ cusp around $\theta = 0°$, it exhibits substantial deviations with $B_c(\theta)$ over a wide $\theta$ range. Instead, a combined formula incorporating the independent 2D and 3D anisotropies provides a better description for the overall $B_c(\theta)$ characteristics (Fig. 2**d**). We therefore ascribe the non-monotonicity of $B_c(\theta)$ to an underlying 3D superconducting anisotropy featuring $B_{c\parallel} < B_{c\perp}$, which should be genetically related to paramagnetic pair-breaking effect considering the large Maki parameter $\alpha$. This subtle 3D anisotropy, hidden in the unpatterned pristine $Nd_{0.8}Sr_{0.2}NiO_2$ films, is unveiled when the nano-pattern modulation weakens the dominant 2D anisotropy. In contrast, the $B_c(\theta)$ curves of the nano-patterned $La_{0.8}Sr_{0.2}NiO_2$ films increase monotonically from $|\theta| = 90°$ to $\theta = 0°$, and remain well-described by the 2D Tinkham model with anisotropy ratio $\gamma$ comparable to that of the unpatterned films (Extended Data Fig. 4). The absence of anisotropy reversal in the nano-patterned La-based films demonstrates that this phenomenon is not simply an extrinsic artifact due to nano-pattern engineering (such as geometric deformation or vortex pinning by the holes), but is fundamentally associated with the intrinsic unconventional superconductivity in Nd-based nickelates.

To directly track the reversal of superconducting anisotropy, the $R_s(T)$ curves under selected in-plane and out-of-plane magnetic fields, and the ratios between them are plotted (Fig. 2**e** to **h**), with the ratios being defined as resistance anisotropy ratio $\gamma_R = R_s(B_\perp = |B|, T)/R_s(B_\parallel = |B|, T)$. At relatively high temperatures where the superconductivity is developing locally in the islands (around $T_1$), the $\gamma_R$ is slightly larger than 1, qualitatively consistent with 2D superconducting anisotropy. At relatively low temperatures where the inter-island phase fluctuations become dominant (below $T_2$), the $\gamma_R$ is reduced even below 1 in the intermediate magnetic field strength regime (e.g., 4 T and 6 T), indicating the emergence of reversed anisotropy. These crossings between $R_S(B_\perp = |B|, T)$ and $R_S(B_\parallel = |B|, T)$ curves directly demonstrate the crossover of dimensionality, where the superconducting state becomes more fragile to the in-plane field rather than out-of-plane field, revealing the unconventional dimensionality of superconductivity.

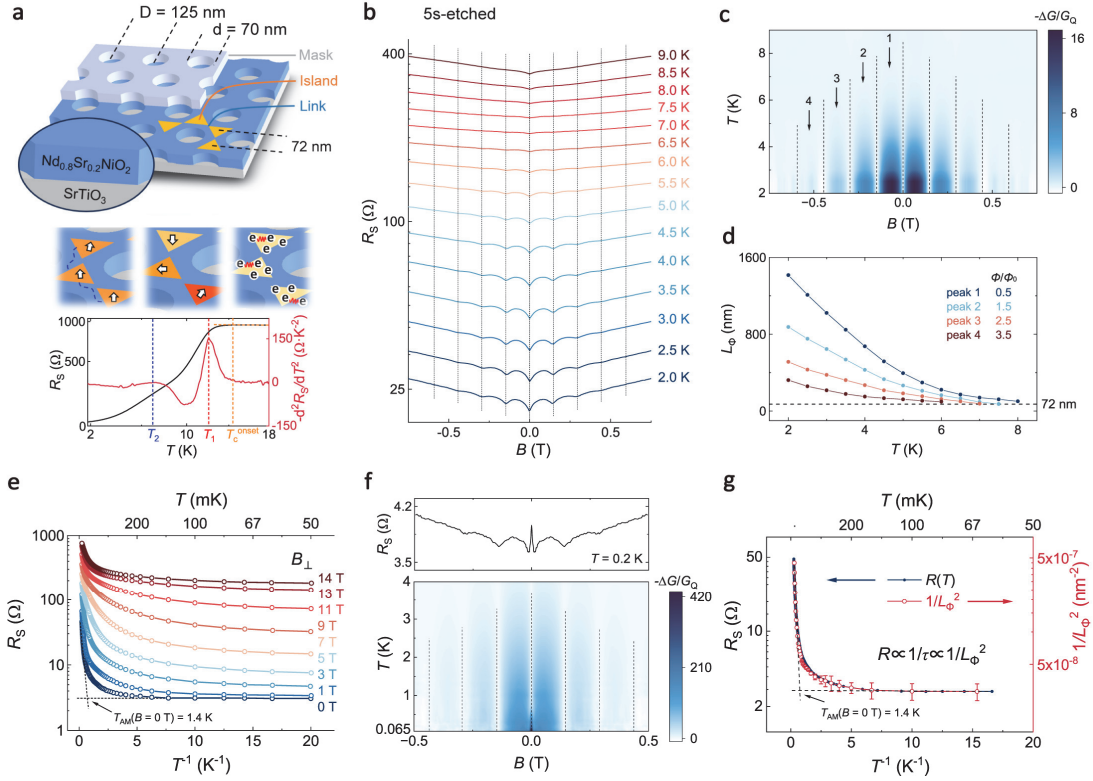

**Fig. 3 | Phase coherence in 5s-etched nano-patterned $Nd_{0.8}Sr_{0.2}NiO_2$ film. a,** Schematic illustrations, $R_S(T)$ curve and second derivative ($-d^2R_S/dT^2$) curve of the two-stage superconducting transition after nano-patterning, marking the onset transition at $T_c^{onset}$, establishment of local superconductivity around $T_1$, and inter-island phase coherence around $T_2$. Yellow triangles illustrate the superconducting islands in JJA. Around $T_c^{onset}$, electrons (marked as $e$) form Cooper pairs in the islands. Around $T_1$, the superconducting phase between the Cooper pairs becomes coherent within the individual islands (represented by arrows), but incoherent among the adjacent islands (different orientations of arrows). Around $T_2$, inter-island phase coherence is mediated by Josephson coupling through the links (blue dashed line). **b,** Magnetoresistance $R_S(B)$ above 2 K, showing $h/2e$ quantum oscillations. **c,** Quantum oscillations of magnetoconductance after the subtraction of backgrounds and the normalization by quantum conductance, $-\Delta G/G_Q$, derived from **b**. **d,** Temperature-dependent phase coherence lengths $L_\Phi$ derived from oscillation amplitudes, as marked in **c**. Black dashed

line corresponds to 72 nm, the center-to-center distance between adjacent superconducting islands, as illustrated in **a**. **e**, Arrhenius plot of $R_S(T)$ down to 50 mK. Black arrow indicates the characteristic temperature for the anomalous metallic state around 1.4 K. **f**, Upper panel shows one representative $R_S(B)$ curve at 0.2 K, highlighting the emergent zero-field magnetoresistance peak below 0.7 K. Lower panel shows the quantum oscillations of magnetoconductance from 4 K to 65 mK, after subtracting the backgrounds. **g**, $R_S(T)$ at $B = 0$ T overlaid with $1/L_\Phi^2$, where $L_\Phi$ is determined by the oscillation amplitudes from **f**. The error bars account for the propagated errors introduced during the subtraction of the magnetoconductance background and the magnetoconductance peak around 0 T. Vertical dashed lines in **b**, **c**, and **f** mark the oscillations with a period of $\Delta B \sim 0.148$ T.

Reflecting the periodicity of the holes array, the magnetoresistance of a 5s-etched nano-patterned $Nd_{0.8}Sr_{0.2}NiO_2$ films shows quantum oscillations with a period of $\Delta B \sim 0.148$ T (Figs. 3**b, c**). This period corresponds to one superconducting flux quantum $\Phi_0 = h/2e = S \cdot \Delta B$ threading an area of one unit cell of the nano-pattern $S = 1.40 \times 10^4$ nm$^2$, where $e$ is the electron charge and $h$ is the Planck's constant. Figure 3**c** presents the background-subtracted magnetoconductance ($\Delta G$) that are further normalized by quantum conductance $G_Q = \frac{4e^2}{h}$, highlighting the fine structures of quantum oscillations. The charge-$2e$ oscillations allow the determination of phase coherence length $L_\Phi$ by $G_{osc} = G_Q \left(\frac{2L_\Phi}{\pi d}\right)^{1.5} \exp\left(-\frac{\pi d}{2L_\Phi}\right)$ with $G_{osc}$ denoting the amplitudes of the oscillations[39]. The derived $L_\Phi$ monotonically decreases with either increasing temperature or winding multiple flux quanta (Fig. 3**d**). The discernible magnetoconductance oscillations yield $L_\Phi$ values consistently exceeding 72 nm (black dashed line in Fig. 3**d**), which corresponds to the center-to-center distance between adjacent superconducting islands (illustrated in Fig. 3**a**), directly linking the emergence of quantum oscillations with the inter-island phase coherence of superconductivity.

In a 2D superconducting system, while the zero-resistance state arises from the macroscopic phase coherence, the loss of quasi-long-range coherence may give rise to the anomalous metallic state, a novel ground state featuring a finite saturating resistance. The dissipation of the anomalous metallic state towards the zero-temperature limit has been previously attributed to the quantum tunneling of vortices (time-domain fluctuations of phase)[40]. To investigate the ground-state characteristics, Figure 3**e** displays the Arrhenius plots ($\lg R_S$ versus $1/T$) of the 5s-etched $Nd_{0.8}Sr_{0.2}NiO_2$ film measured down to 50 mK. With decreasing temperature, the resistance drops rapidly and turns into a saturating plateau below $T^{AM} \sim 1.4$ K, unveiling the emergence of the anomalous metallic state (here, the characteristic temperature for the anomalous metallic state $T^{AM}$ is defined as the crossing point of the extrapolation of resistance drop and saturation). The $h/2e$ quantum oscillations from charge-2e Cooper pairs persist down to the ultralow temperature 65 mK (Fig. 3**f**), demonstrating the bosonic nature of the anomalous metallic state. Consistently, the bosonic nature is further supported by the vanishing Hall response (Fig. S8).

Remarkably, a distinct zero-field magnetoresistance peak is observed at ultralow temperature below 0.7 K, which indicates that the superconductivity is anomalously enhanced by external magnetic fields, signifying the extreme quantum phase fluctuations across the array[45,46] (one representative curve is shown in upper panel of Fig. 3**f**, see also the zero-field peak in lower panel of Fig. 3**f**, more data can be found in Extended Data Fig. 2). Presumably, in our nano-patterned Nd-nickelate array, the strong electronic correlations and local fluctuations driven by the nano-patterning engineering amplify the repulsive Coulomb interactions within the inter-island links. The coherent diffusion of Cooper pairs in this repulsive environment anomalously induces a negative proximity effect ($\Delta < 0$)[47]. This acts as an intrinsic macroscopic phase penalty that severely suppresses the inter-island Josephson coupling at zero field. Applying a small out-of-plane magnetic field introduces an orbital dephasing cutoff that disrupts the interference loop of Cooper pairs, thereby erasing this interaction-induced phase penalty and anomalously restoring the macroscopic conductance.

Interestingly, we find that the resistance saturation is synchronized with $1/L_\Phi^2$ in the low-temperature regime (Fig. 3**g**). This synchronization associates the saturating resistance of the anomalous metallic state and the dephasing of Cooper pairs by $R \propto \frac{1}{\tau_\Phi} \propto \frac{1}{L_\Phi^2}$ with $\tau_\Phi$ denoting the dephasing time[48]. Given the duality between Cooper pairs and vortices, $\tau_\Phi$ could also be described by the dynamics of instanton/anti-instanton (i.e., quantum tunneling of vortices under ohmic dissipation)[40,49]. Our results provide the critical insights into the microscopic origin of anomalous metallic ground state in 2D superconducting systems.

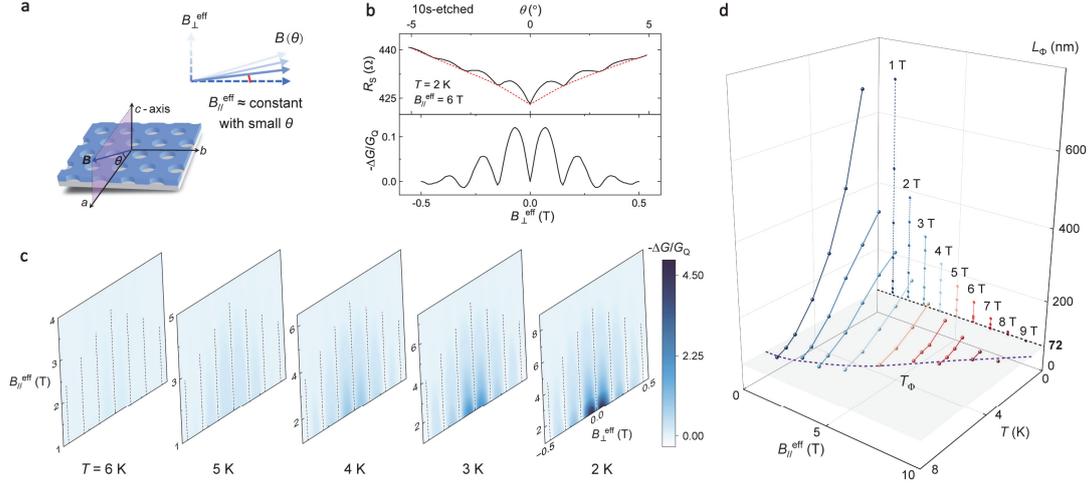

**Fig. 4 | Quantum oscillations in 10s-etched nano-patterned $Nd_{0.8}Sr_{0.2}NiO_2$ film with an effective in-plane field. a**, Illustration for probing quantum oscillations with an effective in-plane field. The magnetic field $B$ rotates in the vicinity of $\theta = 0°$ with a constant strength. With small $\theta$, the effective in-plane component $B_\parallel^{eff} = B \cdot \cos(\theta)$ remain nearly constant, while the effective out-of-plane component $B_\perp^{eff} = B \cdot \sin(\theta)$ contributes to the charge-$2e$ quantum oscillations. **b**, Representative polar angle-dependent sheet resistance $R_S(\theta)$ (upper panel) and background-subtracted magnetoconductance versus $B_\perp^{eff}$ (lower panel) with a magnetic field strength of 6 T at 2 K. **c**, Evolution of magnetoconductance oscillations versus $B_\perp^{eff}$ at different temperatures and $B_\parallel^{eff}$. Vertical dashed lines mark the oscillations with a period of $\Delta B_\perp^{eff} \sim 0.148$ T. **d**, 3D plot of phase coherence length $L_\Phi$ as a function of $T$ and $B_\parallel^{eff}$, alongside the projection onto the $B_\parallel^{eff}$ - $L_\Phi$ plane. Here, $L_\Phi$ is derived from the oscillation amplitude at $\Phi/\Phi_0 = 0.5$. Horizontal plane corresponds to 72 nm, the center-to-center distance between adjacent superconducting islands. Purple dashed line indicates the $B_\parallel^{eff}$-$T$ boundary of inter-island coherence, denoted as $T_\Phi$.

Parallel to thermal fluctuations, the quantum fluctuations driven by an external magnetic field provide a distinct pathway for controlling macroscopic phase coherence. To investigate the inter-island coherence under varying levels of quantum fluctuations, we rotate the magnetic field $B$ in the vicinity around $\theta = 0°$ (*ab*-plane, as schematically illustrated in Fig. 4**a**). With a small $\theta$, the effective in-plane component could be approximately regarded as constant ($B_\parallel^{eff} = |B| \cdot \cos(\theta) \approx |B|$), serving to enhance the quantum fluctuations, while the effective out-of-plane component $B_\perp^{\text{eff}} = |B|\cdot\sin(\theta)$ threads flux and thereby drives the charge-2$e$ quantum oscillations. As shown in Fig. 4**b**, the measured resistance oscillates as magnetic field rotates from -5º to 5º, and the charge-2$e$ quantum oscillations are clearly resolved when the conductance is analyzed as a function of $B_\perp^{\text{eff}}$. As a rigorous control, no discernible oscillation is detected when solely $B_\parallel$ is varied in the absence of $B_\perp$ (Supplementary Information Fig. S14). The amplitudes of quantum oscillations are subject to two independent controllable variables: the quantum fluctuations primarily affected by $B_\parallel^{\text{eff}}$, and the thermal fluctuations by temperature (Fig. 4**c**). Figure 4**d** displays the 3D plot of $L_\Phi$, estimated by the quantum oscillation amplitudes, as a function of both two parameters ($T$ and $B_\parallel^{\text{eff}}$). Consistently, the derived $L_\Phi$ exhibit minimal values above 72 nm (horizontal plane) that corresponds to the center-to-center distance between adjacent superconducting islands, further supporting that the quantum oscillation originates from the coherence across the superconducting islands. The critical boundary of the charge-2$e$ quantum oscillations (purple dashed line) effectively delineates the quantum melting boundary of inter-island coherence in the $B$-$T$ phase diagram of the nano-patterned nickelate film (also plotted as $T_\Phi$ in Fig. 5**e**).

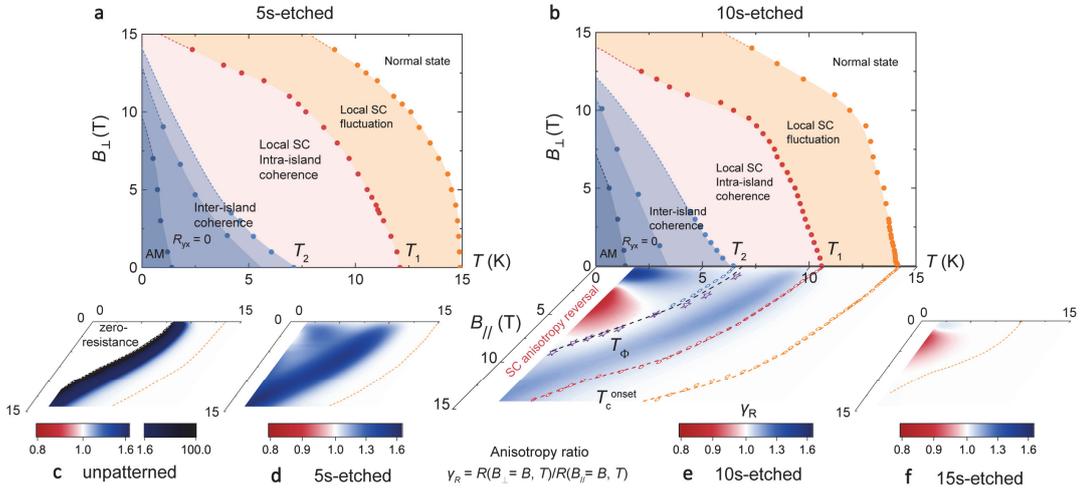

**Fig. 5 | Global phase diagrams for the nano-patterned Nd$_{0.8}$Sr$_{0.2}$NiO$_2$ films. a, b,** $B_\perp$-$T$ phase diagram of 5s-etched (**a**) and 10s-etched (**b**) nano-patterned film. Orange, red, and blue symbols refer to $T_c^{\text{onset}}$, $T_1$, and $T_2$, respectively. With decreasing temperatures, the system successively goes through the normal state, local superconductivity (SC) with fluctuations, local SC with intra-island phase coherence, inter-island phase coherence, zero Hall responses, and the anomalous metallic state

(AM). **c-f,** Evolution of the $B_\parallel$-T phase diagrams upon variation of nano-pattern etching time. The color scales map the anisotropy ratio of resistance $\gamma_R$. Hollow stars in **e** represents the inter-island coherence boundary, determined by the quantum oscillation modulated by $B_\parallel^{\text{eff}}$(denoted as $T_\Phi$). The shift from blue to red color in **e** signifies the reversal of superconducting anisotropy.

Figure 5**a** and **b** represent the $B_\perp$-T phase diagrams of the 5s-etched and 10s-etched Nd$_{0.8}$Sr$_{0.2}$NiO$_2$ films, respectively, which provide a comprehensive and general characterization of how superconductivity develops in the presence of fluctuations enhanced by the nano-patterning. Upon entering the superconducting phase, superconductivity initially emerges within the individual islands with both amplitude and phase fluctuations (local SC fluctuation), the two degrees of freedom of a complex superconducting order parameter. Around $T_1$, the intra-island phase coherence develops to stabilize the local superconductivity. With further decreasing temperature to $T_2$, the inter-island phase coherence progressively stabilizes the global superconductivity, mediated by Josephson coupling across the island array and evidenced by the $h/2e$ quantum oscillations. Then, the Hall coefficient drops to zero, indicative of the bosonic nature of the system with particle-hole symmetry. Approaching zero-temperature limit, the system ends up in the anomalous metallic state, which is dominated by the Cooper pairs and characterized by the quantum tunneling of vortices (time-domain fluctuations of phase)[40].

The $B_\parallel$-T global phase diagrams across the Nd$_{0.8}$Sr$_{0.2}$NiO$_2$ films with different etching time reveal the evolution of unconventional superconductivity driven by nano-pattern modulations, particularly regarding the emergence of anisotropy reversal (Fig. 5**c** to **f**). By mapping the resistance anisotropy ratio $\gamma_R$, defined as $\gamma_R(|B|, T) = R_S(B_\perp = |B|, T)/R_S(B_\parallel = |B|, T)$, the phase diagrams show different patterns of anisotropy that deform progressively. In the unpatterned Nd$_{0.8}$Sr$_{0.2}$NiO$_2$ film, $\gamma_R$ value is relatively large and monotonically increases with decreasing temperatures, which indicates the dominate 2D anisotropy that shadows the underlying 3D characteristics (Fig. 5**c**). In the 5s-etched film, $\gamma_R$ value decreases significantly, and outlines multiple patterns in the phase diagram (Fig. 5**d**). Notably, the unconventional regions with $\gamma_R < 1$, i.e., the reversed anisotropy violating the 2D geometric confinement, emerge in the intermediate magnetic field regime in the 10s-etched film (Fig. 5**e**) and 15s-etched films (Fig. 5**f**), whose occupation in phase diagram expands with increasing etching time. The systematic and dramatic evolution of unconventional superconducting characteristics validate nano-pattern engineering as an effective methodology to investigate and modulate the superconductivity through the strategic manipulation via superconducting phase degree of freedom.

The correlation between the superconducting anisotropy reversal and the establishment of inter-island phase coherence in the 10s-etched film is particularly revealing, as summarized in the $B_{/\!/}$-$T$ phase diagram (Fig. 5e). The phase boundary for inter-island coherence, delineated by charge-$2e$ quantum oscillation under an effective $B_{/\!/}$ (purple stars), encloses the regime where the reversal of superconducting anisotropy emerges ($\gamma_R < 1$, red region). This identifies a highly unconventional regime of superconductivity, where the superconductivity is more fragile to in-plane magnetic field, despite the presence of inter-island phase coherence that should stabilize the effective two-dimensionality. In this regime, the underlying 3D component of anisotropy becomes dominant, overshadowing the typically dominant 2D anisotropy due to geometric confinement, and even driving the out-of-plane critical magnetic fields to exceed the in-plane ones.

We now elucidate the microscopic origin of this anomalous superconducting anisotropy reversal ($B_{c/\!/} < B_{c\perp}$). Notably, pure geometric confinement or generic disorder cannot autonomously account for this phenomenon. As shown in our control experiments (Extended Data Figs. 3-5), the nano-patterned non-magnetic $La_{0.8}Sr_{0.2}NiO_2$ films consistently maintain the conventional 2D anisotropy ($B_{c\perp} < B_{c/\!/}$). Moreover, unpatterned Nd-films merely exhibit a weak 2D anisotropy without reversal ($B_{c\perp} < B_{c/\!/}$). Therefore, this anomalous reversal profoundly unmasks a hidden 3D character intrinsic to the Nd-based system. According to the Mott-Kondo framework[50] and corroborated by the $c$-axis lattice constant-dependent 2D-3D crossover[31,51], this 3D character fundamentally arises from a significantly stronger hybridization between the Ni-$3d_{x^2-y^2}$ and $5d$ orbitals (interstitial $s$) in Nd-nickelates compared to La-nickelates. In our nano-patterned arrays, the artificial geometric constrictions amplify repulsive Coulomb interactions, severely suppressing the macroscopic 2D phase stiffness via intense quantum phase fluctuations[47]. By stripping away this 2D macroscopic phase coherence and concurrently quenching the macroscopic out-of-plane orbital pair-breaking effect, the nano-patterning uniquely exposes the highly hybridized 3D pairing state and its interplay with intrinsic local magnetism.

Once the continuous orbital limit is geometrically quenched, the high-field survival of superconductivity is dictated predominantly by the Pauli paramagnetic limit, bringing the specific role of the rare-earth $4f$ electrons to the forefront. Unlike the itinerant $5d$ orbitals, the Nd-$4f$ electrons reside in deep energy levels and do not directly participate in the pairing, but rather provide a highly anisotropic local magnetic background. This highly anisotropic pair-breaking mechanism plays an elusive role in the pristine Nd-nickelates ($4f^3$), which might represent a singular case bridging the conventional 2D superconducting La-based nickelates ($4f^0$) to the heavily Eu-doped Sm-based nickelates($4f^7$ for $Eu^{2+}$ and $4f^5$ for $Sm^{3+}$) with exotic magnetic field re-entrant superconducting phenomena[52-54] (see detailed discussion in Supplementary Information). Presumably, the nano-patterning may expose the latent magnetic landscape of the $Nd^{3+}$ ground state, uncovering a highly anisotropic effective $g$-factor with an in-plane easy axis ($g_{\parallel} > g_{\perp}$)[55]. In such situation, an applied in-plane field may polarize the Nd moments, generating an internal exchange-Zeeman field that annihilates the singlet pairs, causing a relatively small value of $B_{c\parallel}$. Conversely, an out-of-plane field couples weakly to the $4f$ moments ($g_{\perp}$ is small), preserving the anomalously robust out-of-plane critical field. Moreover, the robust superconductivity under out-of-plane magnetic field is guaranteed by a dual suppression of the orbital pair-breaking effect, specifically the strong Kondo hybridization intrinsically reduces the continuous orbital effect via heavy-fermion mass renormalization, and the periodic nano-patterning completely quenches the orbital effect through geometric truncation. Further investigations are highly inspired to comprehensively understand this abnormal superconductivity, along with its relationship to phase coherence and distinctive emergence in the nickelate superconductors.

Our experimental observations of the superconductivity with enhanced fluctuations in the nano-patterned $Nd_{0.8}Sr_{0.2}NiO_2$ thin films may yield profound implications on the nickelate superconductors and unconventional high-$T_c$ superconducting families. While exhibiting predominant quasi-2D superconducting anisotropy, the pristine $Nd_{0.8}Sr_{0.2}NiO_2$ films demonstrate a complex dimensionality with subtle yet crucial 3D character, in stark contrast to the pristine $La_{0.8}Sr_{0.2}NiO_2$ films with predominant 2D nature. This also establishes a distinct paradigm that fundamentally differs from the nearly isotropic iron-arsenide superconductors[56], thereby offering a distinctive material platform to unravel the mystery of unconventional high-$T_c$ superconductivity. Through the nano-pattern engineering, the phase fluctuations are enhanced in the $Nd_{0.8}Sr_{0.2}NiO_2$ Josephson-junction-array thin films in a controllable manner. Based on this methodology, we provide the universal phase diagrams illustrating the progressive establishment of superconductivity in the presence of significant phase fluctuations, which trace from the emergence of local superconductivity, through the development of macroscopic phase coherence, and eventually to the ground state of anomalous metal. Specifically, the observation of charge-2$e$ Cooper pair quantum oscillations serves to identify the phase coherence between the preformed local superconducting islands, and further provides a quantitative metric for the length scale of phase coherence. The synchronization between the resistance saturation and phase-coherence-length saturation towards zero-temperature limit uncovers a direct link between the ground-state dissipative behaviors and the quantum dynamics of phase fluctuations. It is worth mentioning that the $h/2e$ oscillations can be modulated via an effective in-plane magnetic field, which introduces a new tuning parameter to investigate the quantization phenomenon, and establishes a novel framework to characterize the macroscopic quantum states during quantum melting. Remarkably, with moderate magnetic fields and sufficient etching time, the superconducting anisotropy exhibits an anomalous reversal behavior ($B_{c/\!/} < B_{c\perp}$) in the Nd-based nickelate films, which is driven by the coupling between extreme quantum phase fluctuations and potential local moments. How superconductivity manifests in response to the synergistic modulation of phase fluctuations and magnetic interactions provides critical insights into the pairing mechanism of the nickelate superconducting family. Ultimately, our study pioneers a novel top-down approach to not only understand but also artificially manipulate the unconventional superconductivity and its intricate interplay with the hidden intertwined orders, by utilizing nanostructures as a powerful quantum fluctuation magnifier.

**Methods.**
**Thin film synthesis.**
The 15 nm-thick infinite-layer $R_{0.8}Sr_{0.2}NiO_2$ ($R$ = La and Nd) films are prepared by topochemical reduction of perovskite $R_{0.8}Sr_{0.2}NiO_3$ films (thickness about 14.5~16 nm) without capping layer[36]. The precursor $R_{0.8}Sr_{0.2}NiO_3$ films are deposited on the $TiO_2$-terminated $SrTiO_3$ (001) substrates by pulsed laser deposition (PLD) using 248 nm KrF laser. The substrate temperature is controlled at 620 °C with an oxygen pressure of 200 mTorr during the deposition. A laser fluence of 1 J/cm$^2$ was used to ablate the target and the size of laser spot is about 3 mm$^2$. After deposition, the samples were cooled down in the same oxygen pressure at the rate of 10 °C/min. To perform the topochemical reduction, the samples were sealed in the quartz tube together with 0.1g $CaH_2$. The pressure of the tube is about 0.3 mTorr. Then, the tube was heated up to 300 °C in tube furnace, hold for 2 hours, and naturally cooled down with the ramp rate of 10 °C/min.

**Device fabrication.**

To fabricate the Josephson junction array (JJA)-patterned $R_{0.8}Sr_{0.2}NiO_2$ thin films, we transferred the commercial anodic aluminum oxide (AAO) membrane masks onto the $R_{0.8}Sr_{0.2}NiO_2$ thin films in acetone. The AAO mask has a triangular array of holes with 70 nm in diameter and 125 nm in spacing. Then, the films were etched by the reactive ion etching (RIE) technique with Ar flow (20 sccm) and 200 W radio frequency power (ME-3A, Institute of Microelectronics, Chinese Academy of Sciences). The chamber pressure was kept around 6.0 Pa during etching.

**Transport measurements.**

The transport measurements were carried out in a commercial 16 T Physical Property Measurement System (PPMS-16, Quantum Design) with a dilution refrigerator option, and a 14 T cryogen-free Physical Property Measurement System (PPMS-DynaCool, Quantum Design) with a dilution refrigerator option and a Helium-3 refrigerator option.

**Phenomenological formula for critical magnetic fields**

To characterize the temperature-dependent upper critical fields along in-plane and out-of-plane orientations $B_{c/\!/}(T)$ and $B_{c\perp}(T)$, the phenomenological Ginzburg-Landau (G-L) formula[33] for 2D superconductors (based on the orbital pair-breaking effect under 2D confinement) are applied:

$$B_{c\perp}(T) = \frac{\phi_0}{2\pi\xi_{G-L}^2(0)}\left(1-\frac{T}{T_c}\right) \quad (1)$$

$$B_{c/\!/}(T) = \frac{\sqrt{12}\phi_0}{2\pi\xi_{G-L}(0)d_{sc}}\left(1-\frac{T}{T_c}\right)^{\frac{1}{2}} \quad (2)$$

where $\phi_0$ is the flux quantum, $\xi_{G-L}(0)$ is the zero-temperature G-L coherence length, and $d_{sc}$ is the superconducting thickness.

To analyze the polar angular dependent upper critical fields $B_c(\theta)$, we used the 2D Tinkham model[35]

$$(B_c(\theta)\cos(\theta)/B_{c//})^2 + |B_c(\theta)\sin(\theta)/B_{c\perp}| = 1 \qquad (3)$$

and the 3D Ginzburg-Landau model

$$B_c(\theta) = B_{c//}/(\cos^2(\theta)+\gamma^2\sin^2(\theta))^{1/2} \qquad (4)$$

with $\gamma = B_{c//}/B_{c\perp}$ denoting the superconducting anisotropy ratio.

**Data availability**
All data needed to evaluate the conclusions in the study are present in the paper and/or the Supplementary Information. The data that support the findings of this study are available from the corresponding author upon request.

**Acknowledgements**
We thank Meixuan Li for the assistance, and Jun Ge for the discussions. This work was financially supported by the National Natural Science Foundation of China [Grant No. 12488201 (J.W.)], the Innovation Program for Quantum Science and Technology [2021ZD0302403 (J.W.)], the Natural Science Foundation of China [Grant No. 92565303 (Y.L.)], the National Key Research and Development Program of China [No. 2023YFA1406500 (Y.L.), No. 2022YFA1403103 (Y.L.)], the Natural Science Foundation of China [Grant No. 12174442 (Y.L.)], the Fundamental Research Funds for the Central Universities and the Research Funds of Renmin University of China [No. 22XNKJ20 (Y.L.)], National Natural Science Foundation of China [Grant No. 52525208 (L.Q.)], Sichuan Science and Technology Program [Grant No. 2024ZYD0164 (L.Q.)], the Key Research and Development Program from the Ministry of Science and Technology [Grant No. 2023YFA1406301 (L.Q.)], the National Natural Science Foundation of China [Grant No. 12474136 (Y.-F.Y.)], the China Postdoctoral Science Foundation [Grant No. 2024M760063 (H.J)],

**Author contributions**
J.W. conceived the project. H.J., Z.X., Z.C. and Y.L. performed the transport measurements and analyzed the data under the guidance of J.W.. X.F. and M.X.

performed the thin film growth under the guidance of L.Q.. G.-M.Z., Y.-F.Y. and H.L. contributed to the theoretical explanations. H.J., Z.C., H.L., Y.L. and J.W. wrote the manuscript with input from all other authors. H.J., Z.X. and X.F. contributed equally to this work.

**Competing interests**

The authors declare no competing interests.

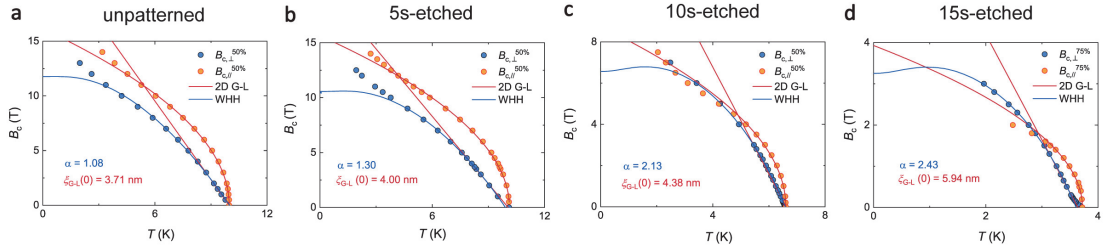

**Extended Data Fig. 1 | $B_c(T)$ curves in the unpatterned and nano-patterned $Nd_{0.8}Sr_{0.2}NiO_2$ films**. $B_c(T)$ curves in the unpatterned (**a**), 5s-etched (**b**), 10s-etched (**c**) and 15s-etched (**d**) $Nd_{0.8}Sr_{0.2}NiO_2$ film. Red solid lines are the theoretical fits using 2D Ginzburg-Landau formula. Blue solid lines represent the WHH fits. Maki parameter $\alpha$ increases monotonically as etching time increases.

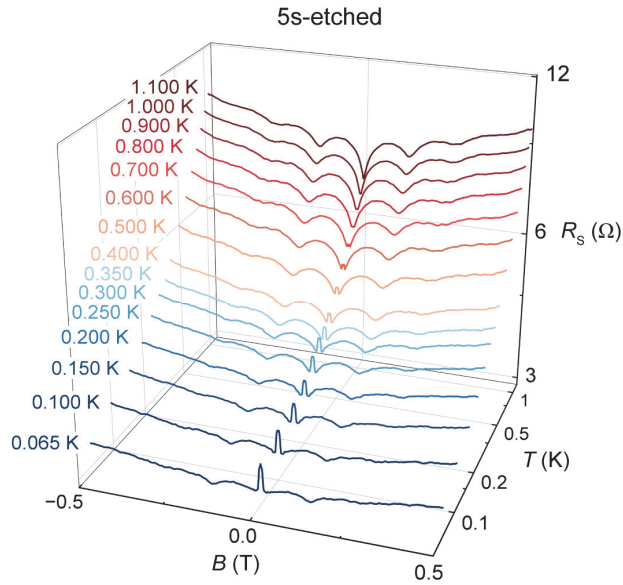

**Extended Data Fig. 2 | Zero-field magnetoresistance peak in the 5s-etched $Nd_{0.8}Sr_{0.2}NiO_2$ film**. $R_S(B)$ curves at different temperatures below 1.1 K in the 5s-etched $Nd_{0.8}Sr_{0.2}NiO_2$ film. Zero-field magnetoresistance peaks, indicative of extreme quantum phase fluctuations, are observed below approximately 0.7 K.

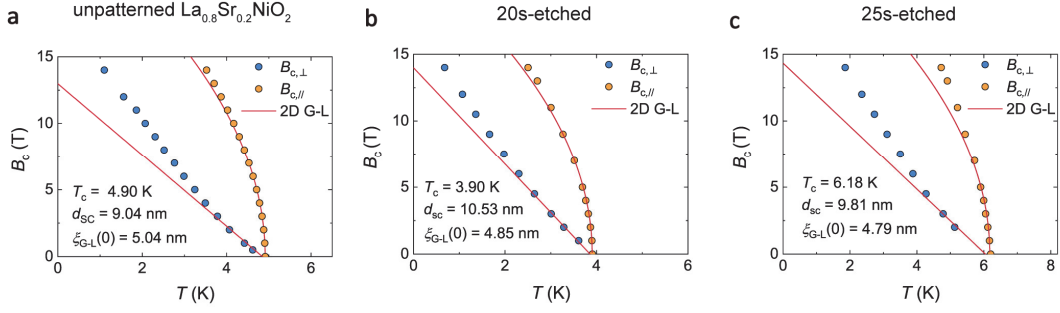

**Extended Data Fig. 3 | $B_c(T)$ curves in the unpatterned and nano-patterned La$_{0.8}$Sr$_{0.2}$NiO$_2$ films**. $B_c(T)$ curves in the unpatterned (**a**), 20s-etched (**b**), and 25s-etched (**c**) La$_{0.8}$Sr$_{0.2}$NiO$_2$ film. Red solid lines are the theoretical fits using 2D Ginzburg-Landau formula.

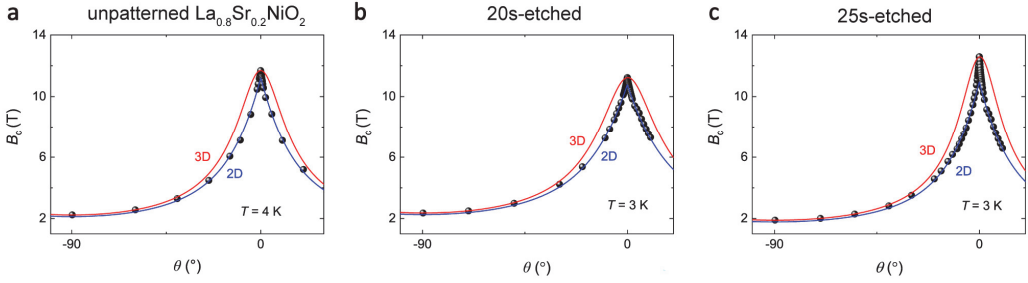

**Extended Data Fig. 4 | $B_c(\theta)$ curves in the unpatterned and nano-patterned La$_{0.8}$Sr$_{0.2}$NiO$_2$ films**. $B_c(\theta)$ curves of the unpatterned La$_{0.8}$Sr$_{0.2}$NiO$_2$ film at $T = 4$ K (**a**), the 20s-etched La$_{0.8}$Sr$_{0.2}$NiO$_2$ film at $T = 3$ K (**b**), and the 25s-etched La$_{0.8}$Sr$_{0.2}$NiO$_2$ film at $T = 3$ K (**c**). Blue and red curves are the theoretical fits using 2D Tinkham model and 3D G-L model, respectively.

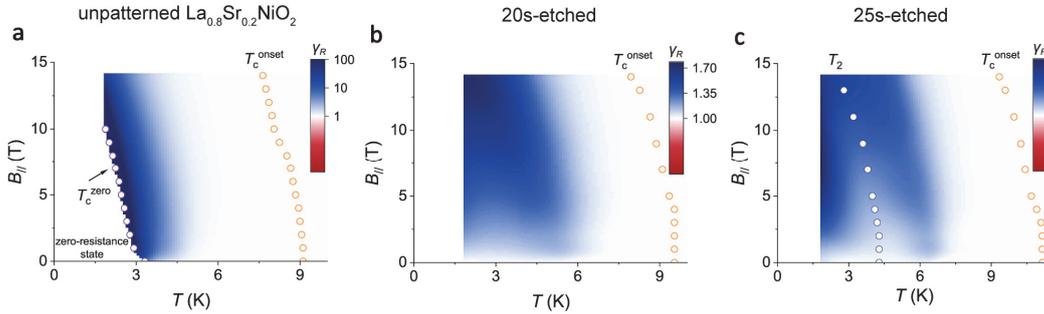

**Extended Data Fig. 5 | $B_{/\!/}$ versus $T$ phase diagrams in the unpatterned and nano-patterned La$_{0.8}$Sr$_{0.2}$NiO$_2$ films.** $B_{/\!/}$-$T$ phase diagrams in the unpatterned (**a**), 20s-etched (**b**), 25s-etched (**c**) films. The color scale denotes the anisotropy ratio of resistance $\gamma_R$. In the nano-patterned La$_{0.8}$Sr$_{0.2}$NiO$_2$ films, $\gamma_R$ is always larger than 1 in

the superconducting phases, which show no reversal of superconducting anisotropy, differing from the phenomenon in the nano-patterned $Nd_{0.8}Sr_{0.2}NiO_2$ films.

# Supplementary Information for
# "Emergent quantum phenomena via phase-coherence engineering in infinite-layer nickelate superconductors"

**I. Possible impact of local moments in infinite-layer nickelate films with different rare earth elements.**

In the Eu-system, the half-filled Eu$^{2+}$ ($4f^7$) local moments possess zero orbital angular momentum ($L = 0$) and a positive de Gennes factor $(g_J - 1) > 0$ [57]. Combined with an antiferromagnetic exchange coupling to the conduction electrons, this produces an internal exchange field that perfectly opposes the external applied field, in which the Jaccarino-Peter (J-P) compensation effect leads to robust field-induced re-entrant superconductivity. In Nd$^{3+}$ ($4f^3$), however, the large orbital angular momentum ($L = 6$) yields a negative de Gennes factor $(g_J - 1) < 0$. This fundamental sign reversal causes the internal exchange field to align parallel to the external magnetic field, drastically amplifying the Zeeman pair-breaking effect rather than compensating it. Furthermore, under the extreme tetragonal crystalline electric field, the Nd$^{3+}$ ground state exhibits an in-plane easy axis with a highly anisotropic effective $g$-factor ($g_{\parallel} > g_{\perp}$)[55]. Therefore, an applied in-plane field may polarize the Nd moments, generates an internal exchange-Zeeman field, leading to relatively small $B_{c\parallel}$. Conversely, an out-of-plane field couples weakly to the $4f$ moments (small $g_{\perp}$), leading to a robust out-of-plane critical field. Moreover, the robust superconductivity under out-of-plane magnetic field is guaranteed by a dual suppression of the orbital pair-breaking effect, originating from the strong Kondo hybridization-induced heavy-fermion mass renormalization and the periodic nano-pattern-induced geometric truncation.

## II. Comparison across un-patterned and nano-patterned Nd$_{0.8}$Sr$_{0.2}$NiO$_2$ films.

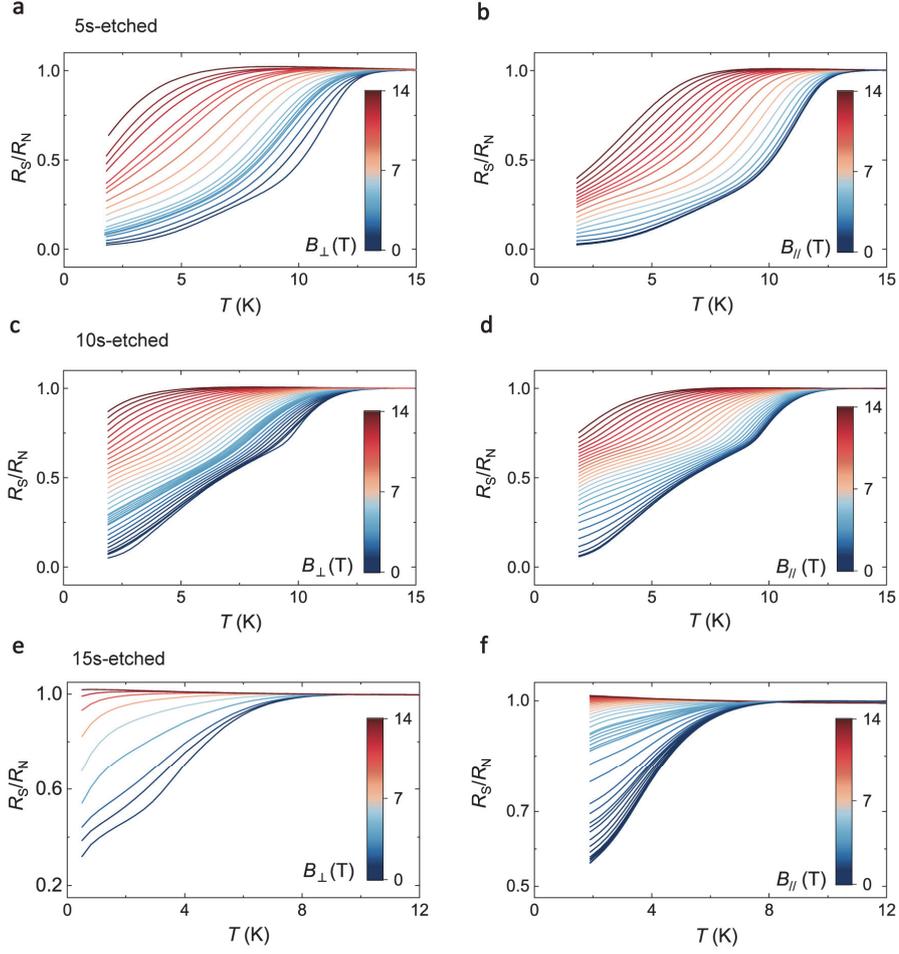

**Fig. S1. Modulation of superconducting transitions in the nano-patterned Nd$_{0.8}$Sr$_{0.2}$NiO$_2$ films.** Temperature-dependent sheet resistance $R_S(T)$ curves normalized by $R_N$ for the 5s-etched (**a** and **b**), 10s-etched (**c** and **d**), and 15s-etched (**e** and **f**) nano-patterned Nd$_{0.8}$Sr$_{0.2}$NiO$_2$ films under different out-of-plane (left column) and in-plane (right column) magnetic fields. $R_S(T)$ curves of 15s-etched film under $B_\perp$ are measured by Helium-3 option down to 0.5 K, and the rest are measured down to 2 K.

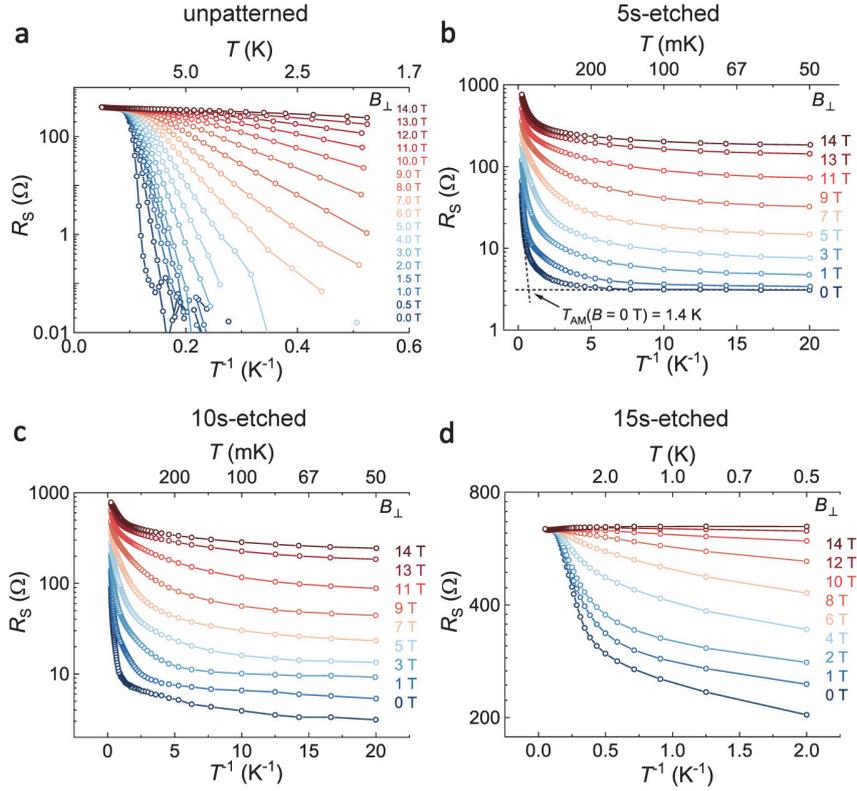

**Fig. S2. Modulation of ground state behaviors in the unpatterned and nano-patterned Nd$_{0.8}$Sr$_{0.2}$NiO$_2$ films.** Arrhenius plot of $R_S(T)$ curves for unpatterned (**a**), 5s-etched (**b**), 10s-etched (**c**) and 15s-etched (**d**) films under out-of-plane magnetic fields. $R_S(T)$ curves of the unpatterned film are measured down to 1.8 K in a PPMS system, which reach zero within the resolution below $B_\perp$ of 5 T (**a**), indicating the zero-resistance superconducting ground state of the unpatterned film. $R_S(T)$ curves of the 5s-etched and 10s-etched films are measured down to 50 mK in the dilution refrigerator option of PPMS, which show saturating behaviors approaching zero temperature (**b** and **c**), indicating the ground states of these two nano-patterned films are anomalous metallic states. $R_S(T)$ curves of the 15s-etched film are measured down to 0.5 K in a helium-3 option of PPMS.

# III. Additional information for unpatterned Nd$_{0.8}$Sr$_{0.2}$NiO$_2$ film.

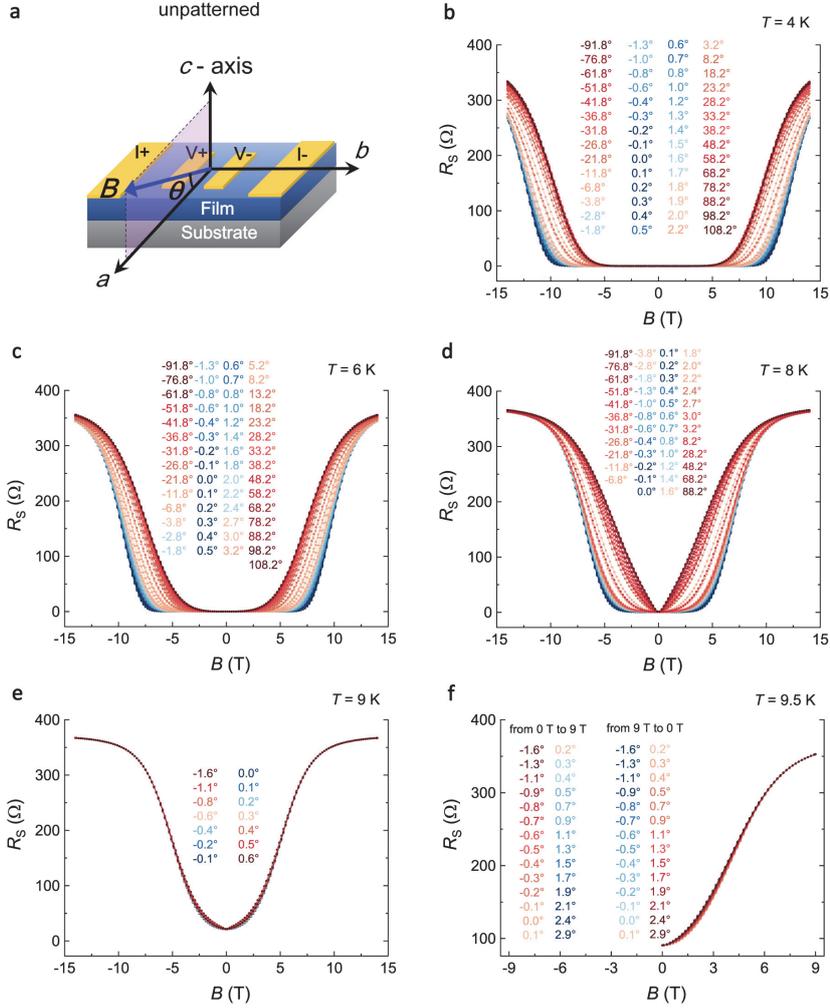

**Fig. S3.** $R_S(B)$ curves at different polar angle $\theta$ for the determination of $B_c(\theta)$. **a,** Schematic of the measurement configuration. **b-f,** $R_S(B)$ curves with magnetic field applied along different polar angle $\theta$ at $T = 4$ K (**b**), 6 K (**c**), 8 K (**d**), 9 K (**e**), and 9.5 K (**f**).

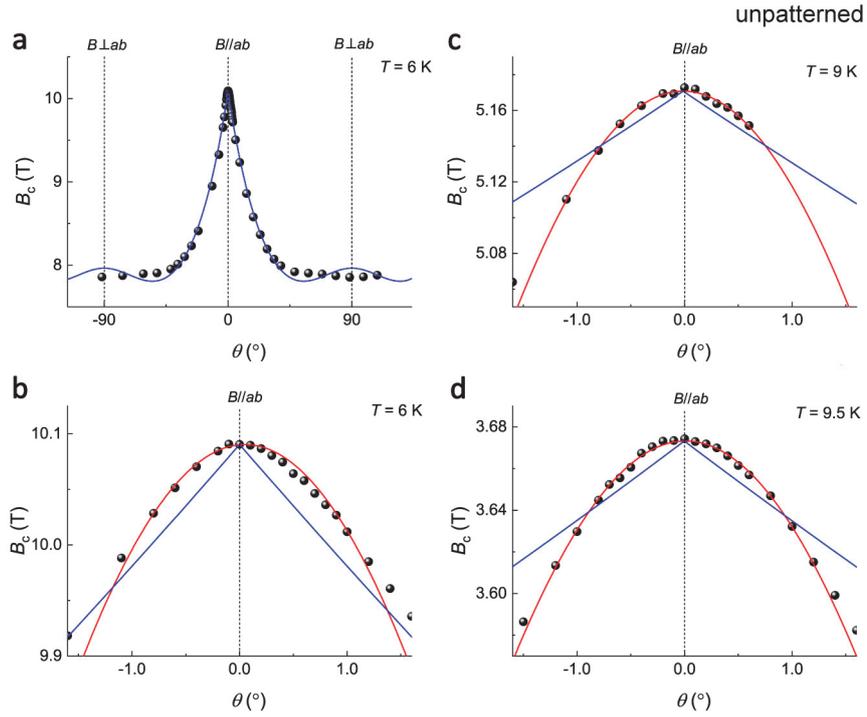

**Fig. S4.** $B_c(\theta)$ **curves at different temperature in the unpatterned $Nd_{0.8}Sr_{0.2}NiO_2$ film**. **a-d,** $B_c(\theta)$ curves at $T = 6$ K (**a** and **b**), 9 K (**c**), and 9.5 K (**d**). Blue and red solid lines are the theoretical curves using 2D Tinkham model and 3D G-L model, respectively. $B_c(\theta)$ curves at different temperatures show the dominant quasi-2D nature and a subtle 3D feature within a narrow $\theta$ range around the in-plane direction.

## IV. Additional information for 5s-etched nano-patterned $Nd_{0.8}Sr_{0.2}NiO_2$ films.

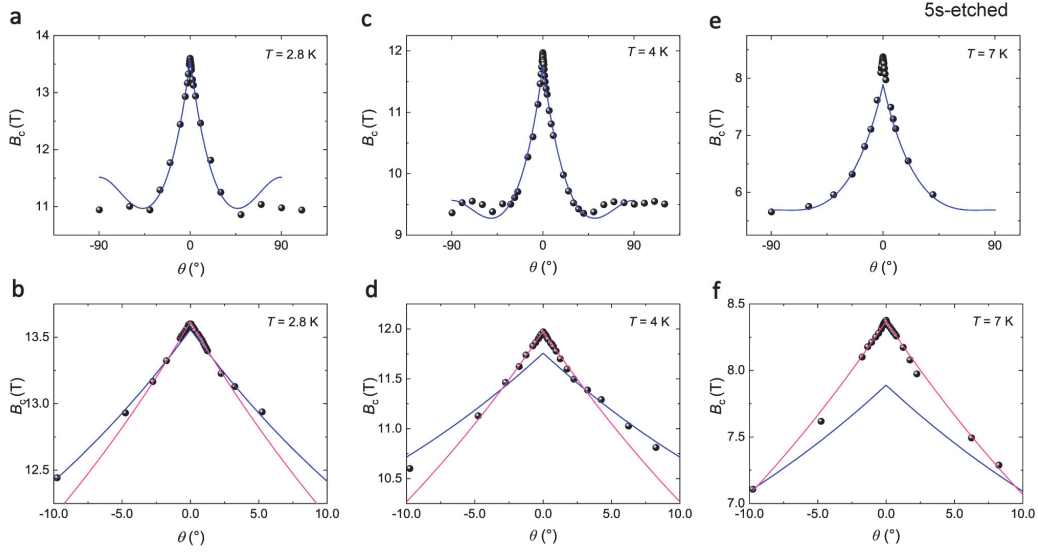

**Fig. S5. $B_c(\theta)$ curves at different temperature in the 5s-etched $Nd_{0.8}Sr_{0.2}NiO_2$ film.** **a-f,** $B_c(\theta)$ curves at $T = 2.8$ K **(a and b)**, 4 K **(c and d)**, and 7 K **(e and f)** of the 5s-etched $Nd_{0.8}Sr_{0.2}NiO_2$ film. Blue and pink solid lines are the theoretical curves using 2D Tinkham model with different parameters. The lower panels **(b, d,** and **f)** correspond to the enlarged view of the upper ones **(a, b,** and **c)**, respectively. The $B_c(\theta)$ curves of the 5s-etched film exhibit dominant quasi-2D nature, and show additional sharp enhancements within a narrow $\theta$ range around the in-plane direction. This additional enhancements are captured by additional 2D fits, reminiscent of the orbital effect-induced finite-momentum pairing state[58].

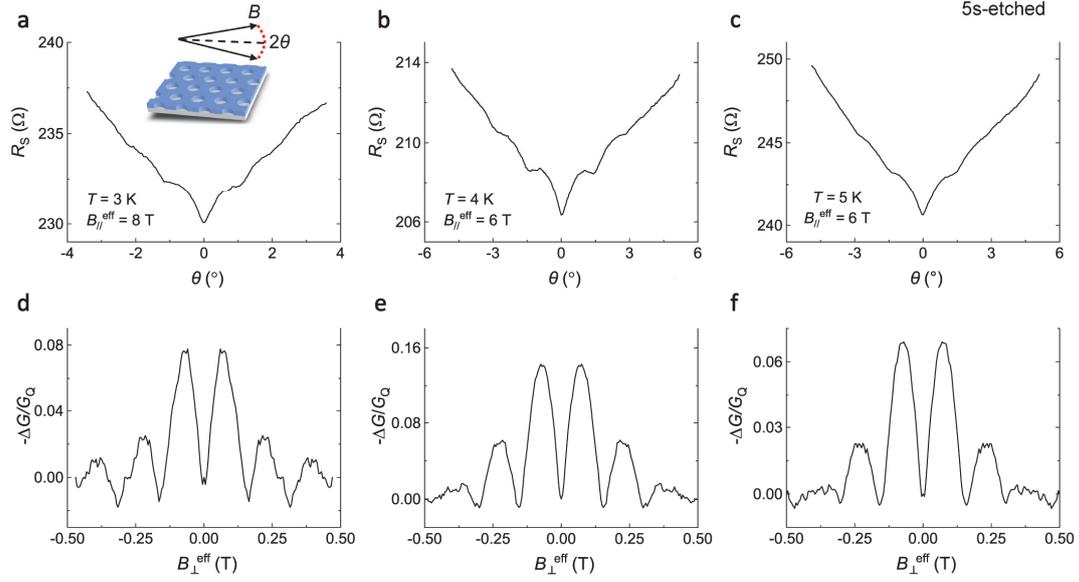

**Fig. S6. Charge-2*e* quantum oscillations with an effective modulation of $B_\parallel^{\text{eff}}$ in the 5s-etched Nd$_{0.8}$Sr$_{0.2}$NiO$_2$ films. a-c**, $R_s(\theta)$ curves in the vicinity of $\theta = 0°$ at different temperatures and magnetic field strengths. **d-f**, Background-subtracted magnetoconductance as a function of $B_\perp^{\text{eff}}$, derived from **a-c**, respectively.

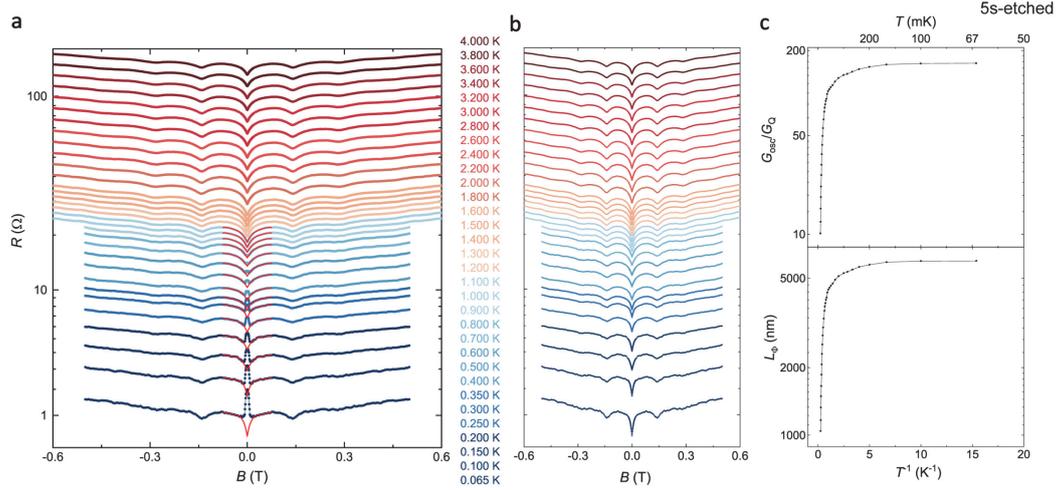

**Fig. S7. Charge-2e quantum oscillations of 5s-etched Nd$_{0.8}$Sr$_{0.2}$NiO$_2$ film in the low temperatures. a**, $R(B)$ curves of 5s-etched Nd$_{0.8}$Sr$_{0.2}$NiO$_2$ film from 0.065 K to 4 K measured in the dilution refrigerator option of PPMS. Below 0.7 K, the $R(B)$ curves show peaks around $B = 0$ T, which may be due to the strong fluctuations[47]. The oscillation structures around $B = 0$ T below 1.1 K are fitted with a polynomial function (red lines). **b**, $R(B)$ curves from **a**, where the peak around $B = 0$ T are replaced by the polynomial fits, in order to subtract the $R(B)$ background and then to derive $G_{osc}$ and $L_\Phi$. The curves in **a** and **b** are shifted for clarity. **c**, Temperature-dependence of the normalized oscillation amplitudes $G_{osc}/G_Q$ (upper panel) and phase coherence length $L_\Phi$ (lower panel) derived from **b**, which saturate at low temperature regimes.

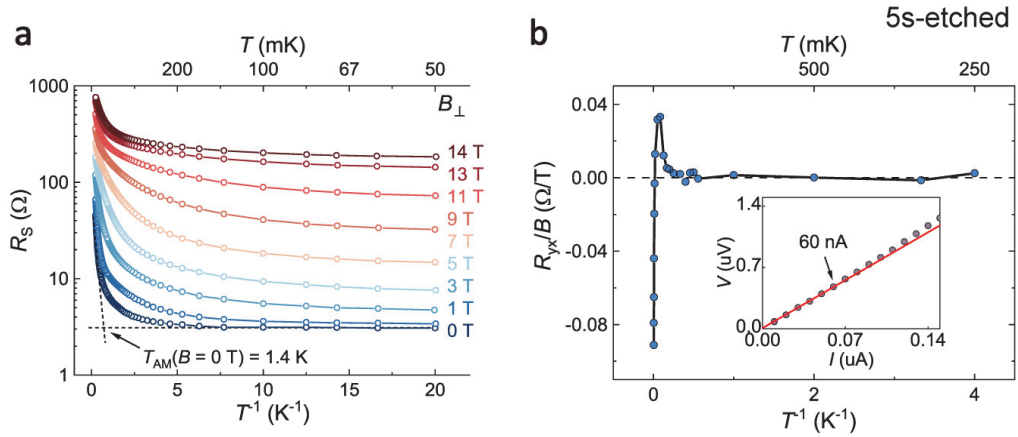

**Fig. S8. Anomalous metallic state and its characteristic in the 5s-etched $Nd_{0.8}Sr_{0.2}NiO_2$ film. a**, Arrhenius plot of $R_S(T)$ under different perpendicular magnetic fields. **b**, Temperature dependent Hall coefficient $R_{yx}/B$, which goes to zero within the measurement resolution below 4 K, as indicated by the black dash line. Inset: *I-V* curve at 0.1 K showing a linear behavior below 60 nA. The excitation current for ultralow-temperature measurement is 50 nA within the ohmic regime.

# V. Additional information for the 10s-etched nano-patterned Nd$_{0.8}$Sr$_{0.2}$NiO$_2$ film.

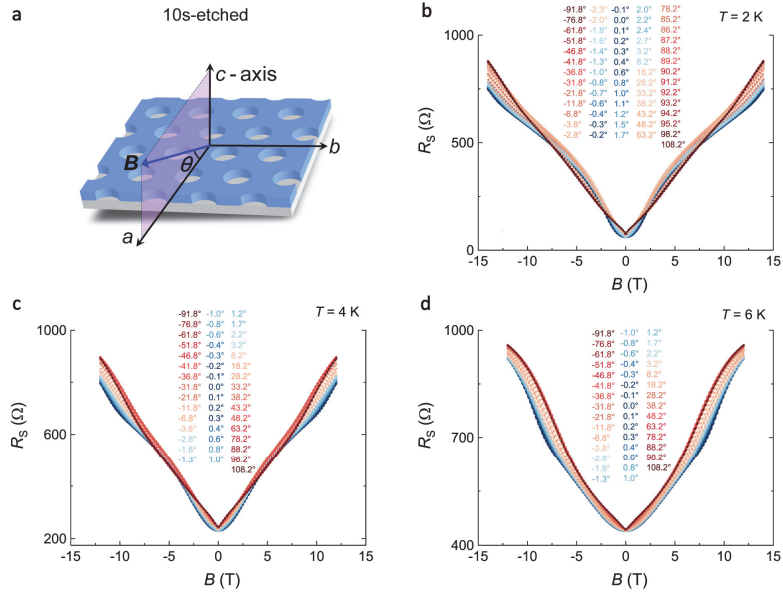

**Fig. S9. $R_S(B)$ curves at different polar angle $\theta$ for the determination of $B_c(\theta)$. a,** Schematic of the measurement configuration. **b-d,** $R_S(B)$ curves with magnetic field applied along different polar angle $\theta$ at $T = 2$ K (**b**), 4 K (**c**), and 6 K (**d**).

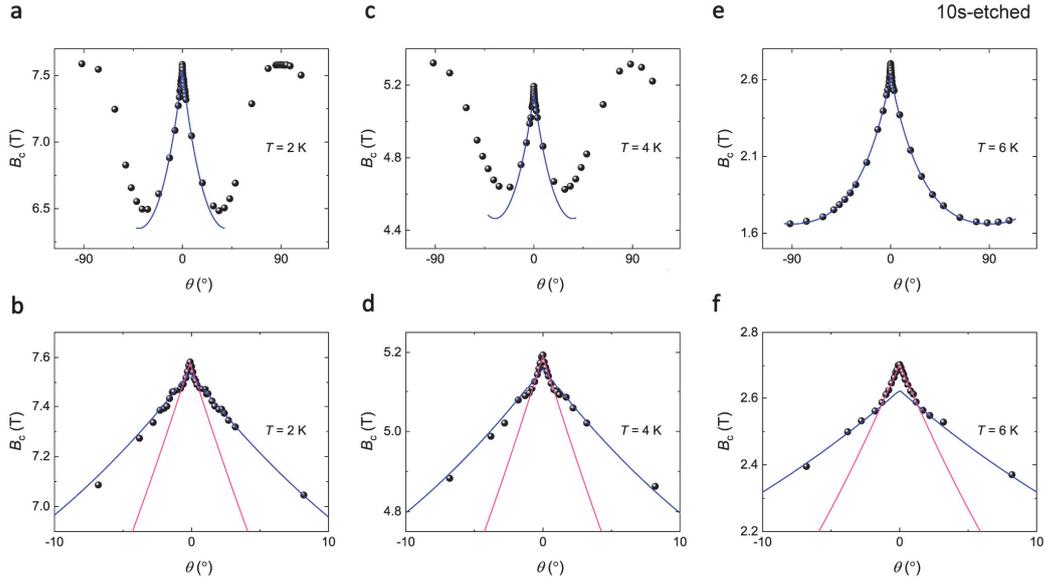

**Fig. S10. $B_c(\theta)$ curves at different temperature in the 10s-etched $Nd_{0.8}Sr_{0.2}NiO_2$ film. a-f,** $B_c(\theta)$ curves at $T$ = 2 K (**a** and **b**), 4 K (**c** and **d**), and 6 K (**e** and **f**) of the 10s-etched $Nd_{0.8}Sr_{0.2}NiO_2$ film. Blue and pink solid lines are the theoretical curves using 2D Tinkham model with different parameters. The lower panels (**b, d,** and **f**) correspond to the enlarged view of the upper ones (**a, b,** and **c**), respectively. The $B_c(\theta)$ curves of the 10s-etched film exhibit quasi-2D characteristics around the in-plane direction, and show additional sharp enhancements within a narrow $\theta$ range. This additional enhancements are captured by additional 2D fits, reminiscent of the orbital effect-induced finite-momentum pairing state[58].

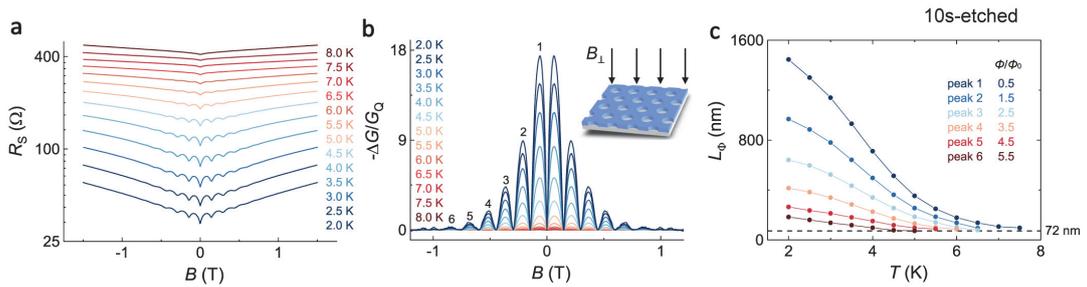

**Fig. S11. Charge-2$e$ quantum oscillations in the 10s-etched $Nd_{0.8}Sr_{0.2}NiO_2$ films above 2 K. a, b,** Quantum oscillations of magnetoresistance (**a**) and magnetoconductance (**b**) after the subtraction of backgrounds from **a** at different temperatures. Magnetic fields are applied perpendicular to the nano-patterned films, as schematically shown in **b**. **c,** Temperature-dependent phase coherence lengths $L_\Phi$ derived from different oscillation peaks in **b**. Black dashed line corresponds to 72 nm, the center-to-center distance between adjacent superconducting islands.

It is noted that the quantum oscillations are unexpectedly robust, considering that $Nd_{0.8}Sr_{0.2}NiO_2$ films are in the dirty limit[27,59]. Up to 12 peaks of the oscillations could be distinguished. The extracted magnetoconductance oscillation at 2 K are 17 times larger than quantum conductance, pointing to a relatively long $L_\Phi$ over 1.4 μm. These results imply the robust phase coherence of Cooper pairs in $Nd_{0.8}Sr_{0.2}NiO_2$ films against the scattering and interference after winding multiple $\Phi_0$.

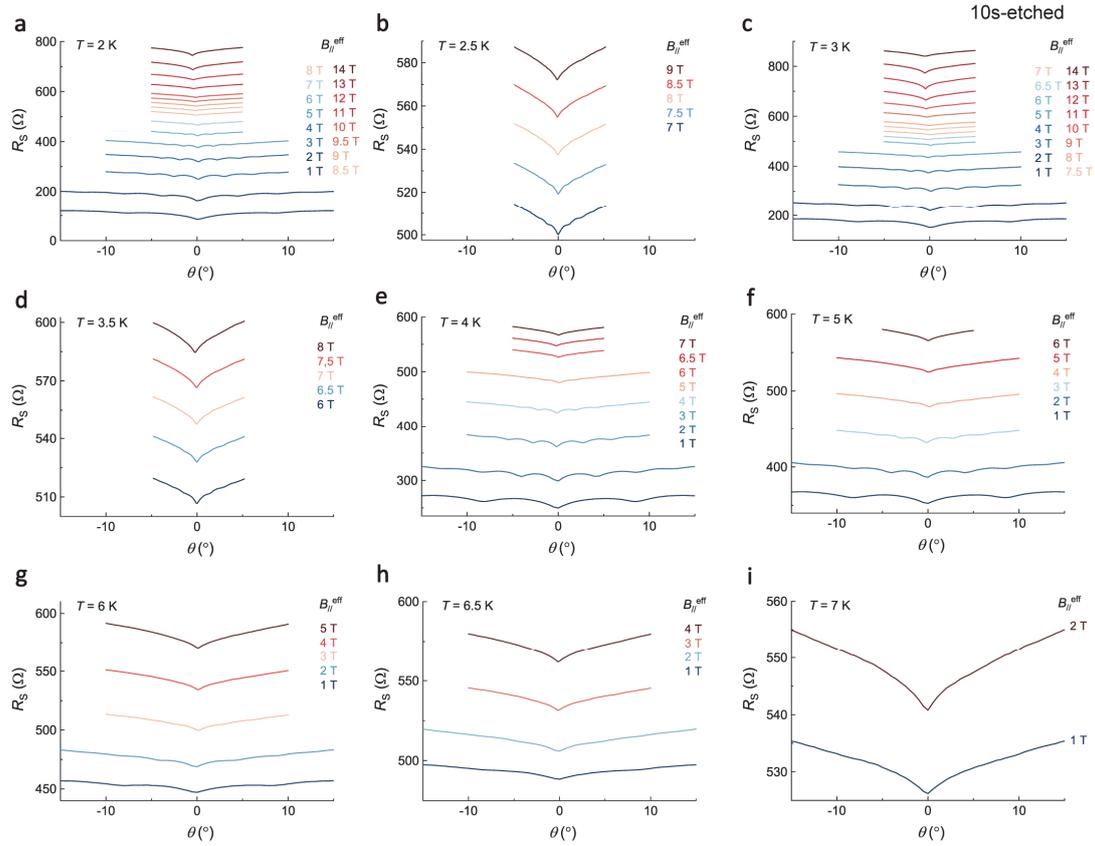

**Fig. S12. $R_S(\theta)$ curves at different temperatures and $B_\parallel^{eff}$. a-i,** $R_S(\theta)$ curves measured with magnetic field rotating in the vicinity of $\theta = 0°$, at $T = 2$ K **(a)**, 2.5 K **(b)**, 3 K **(c)**, 3.5 K **(d)**, 4 K **(e),** 5 K **(f)**, 6 K **(g)**, 6.5 K **(h)**, and 7 K **(i)**.

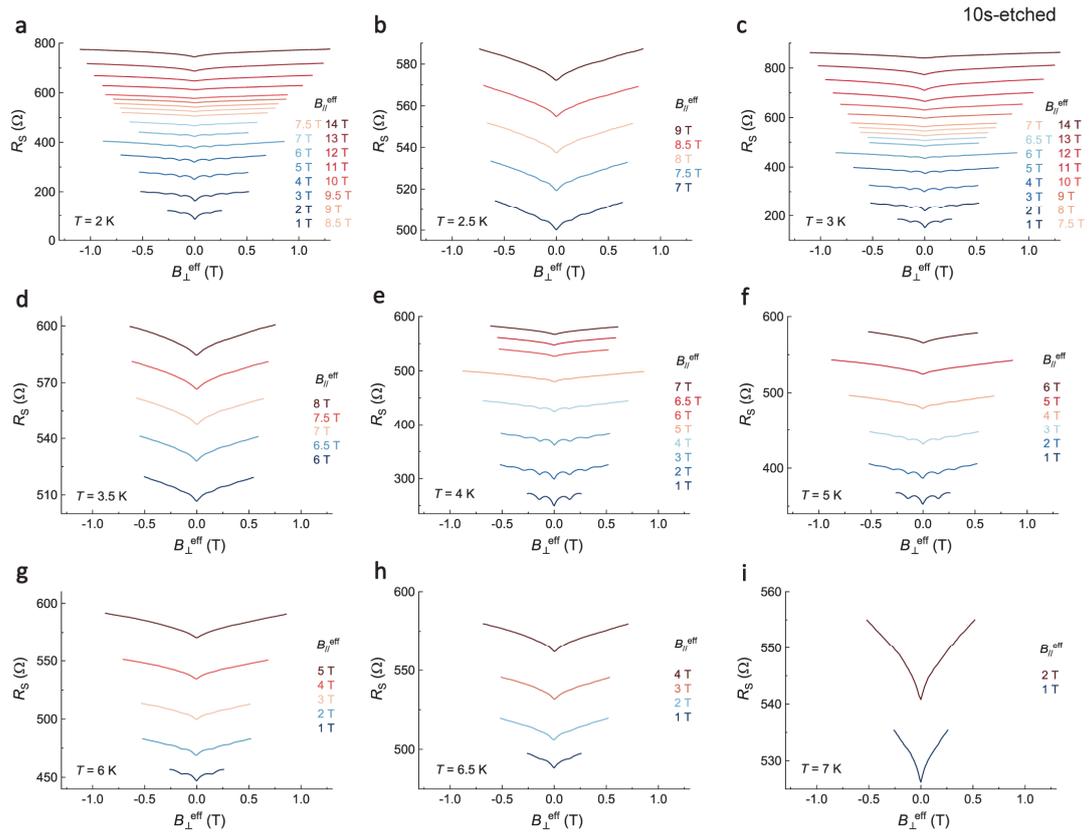

**Fig. S13.** $R_S$ **as a function of** $B_\perp^{eff}$ **at different temperatures and** $B_{//}^{eff}$. **a-i,** $R_S$ versus $B_\perp^{eff}$ curves at $T$ = 2 K **(a)**, 2.5 K **(b)**, 3 K **(c)**, 3.5 K **(d)**, 4 K **(e),** 5 K **(f)**, 6 K **(g)**, 6.5 K **(h)**, and 7 K **(i)**. The data are converted from Fig. S12.

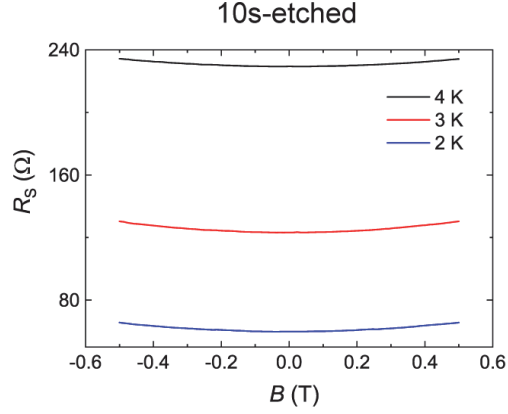

**Fig. S14. $R_S(B)$ curves under in-plane magnetic fields showing no quantum oscillations.** $R_S(B)$ curves with swept in-plane magnetic fields and without out-of-plane fields. No discernable quantum oscillation is observed within the measurement resolution. These results indicate that $h/2e$ quantum oscillations originate from the out-of-plane magnetic field-induced flux variations.

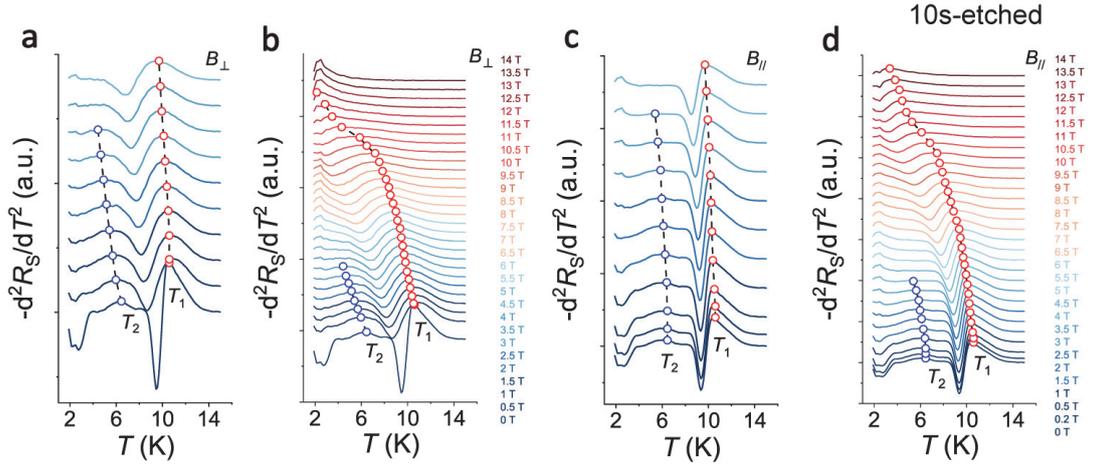

**Fig. S15. Second derivative ($-d^2R_S/dT^2$) of $R_S$ ($T$) for the determination of $T_1$ and $T_2$ in the 10s-etched $Nd_{0.8}Sr_{0.2}NiO_2$ film. a-d,** $-d^2R_S/dT^2$ in the 10s-etched $Nd_{0.8}Sr_{0.2}NiO_2$ film under out-of-plane (**a** and **b**) and in-plane (**c** and **d**) magnetic fields. **a** and **c** focus on the curves under relatively small magnetic fields of **b** and **d**, respectively. Red and blue hollow dots denote the local maxima peaks in the $-d^2R_S/dT^2$, defining the characteristic temperatures of local superconductivity temperature $T_1$ and inter-island phase coherence $T_2$, respectively. Here, the curves are vertically shifted for clarity.

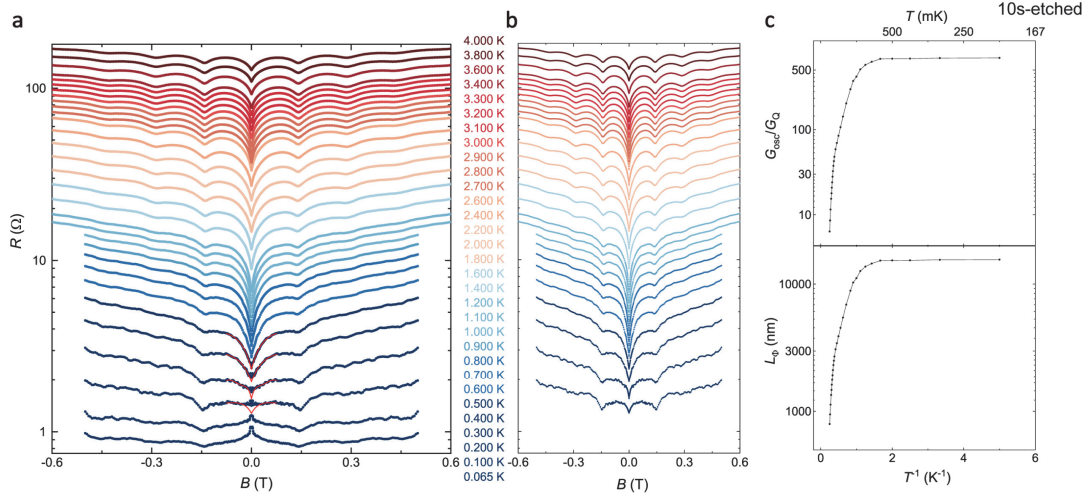

**Fig. S16. Charge-2$e$ quantum oscillations of 10s-etched Nd$_{0.8}$Sr$_{0.2}$NiO$_2$ film in the low temperatures. a**, $R(B)$ curves of 10s-etched Nd$_{0.8}$Sr$_{0.2}$NiO$_2$ film from 0.065 K to 4 K measured in the dilution refrigerator option of PPMS. Below 0.5 K, the $R(B)$ curves show peaks around $B = 0$ T, which may be due to the strong fluctuations[47]. The oscillation structures around $B = 0$ T below 0.5 K are fitted with a polynomial function (red lines). **b**, $R(B)$ curves from **a**, where the peaks around $B = 0$ T are replaced by the polynomial fits, in order to subtract the $R(B)$ background and then to derive $G_{osc}$ and $L_\Phi$. The curves in **a** and **b** are shifted for clarity. **c**, Temperature-dependence of the normalized oscillation amplitudes $G_{osc}/G_Q$ (upper panel) and phase coherence length $L_\Phi$ (lower panel) derived from **b**, which saturate at low temperature regimes.

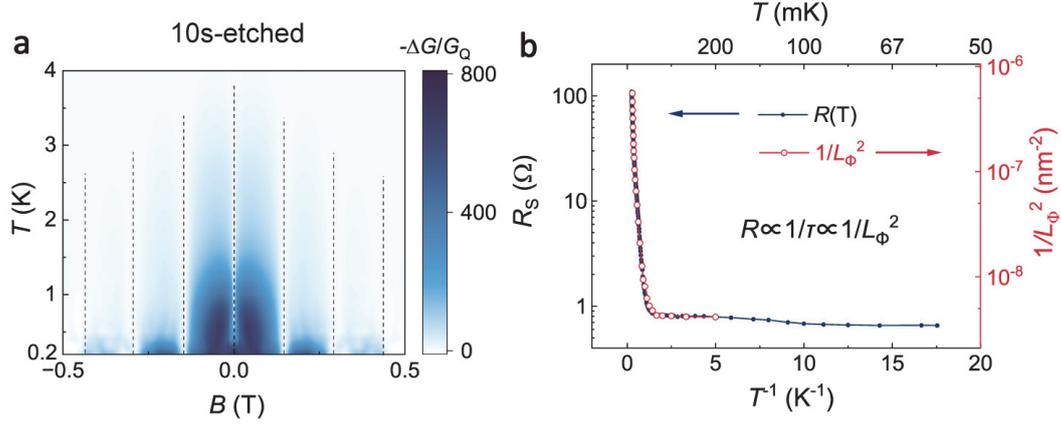

**Fig. S17. Phase coherence length in 10s-etched $Nd_{0.8}Sr_{0.2}NiO_2$ film. a**, Quantum oscillations of magnetoconductance down to 200 mK after the subtraction of the backgrounds. **b**, $R_S(T)$ at $B = 0$ T overlaid with $1/L_\Phi^2$, where $L_\Phi$ is determined by the oscillation amplitudes from **a**.

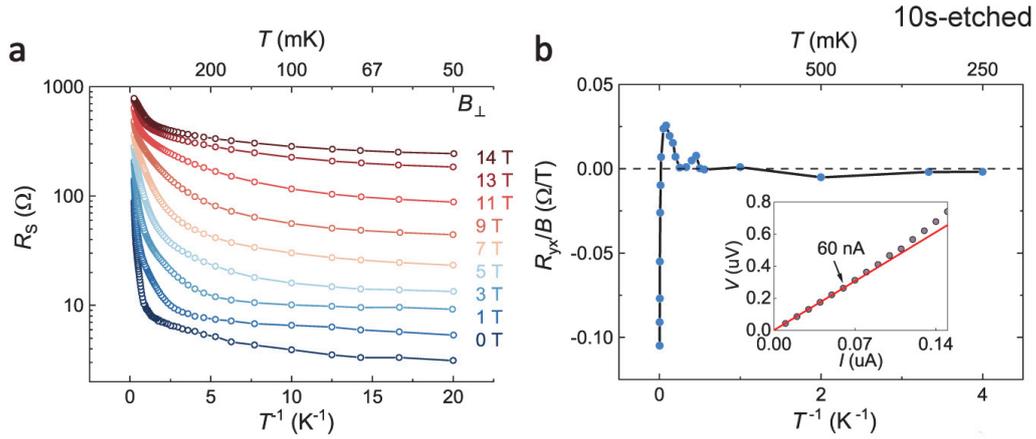

**Fig. S18. Anomalous metallic state and its characteristic in the 10s-etched $Nd_{0.8}Sr_{0.2}NiO_2$ film. a**, Arrhenius plot of $R_S(T)$ under different perpendicular magnetic fields. **b**, Temperature dependent Hall coefficient $R_{yx}/B$, which goes to zero within the measurement resolution below 4 K, as indicated by the black dash line. Inset: $I$-$V$ curve at 0.1 K showing a linear behavior below 60 nA. The excitation current for ultralow-temperature measurement is 50 nA within the ohmic regime.

# VI. Additional information for the 15s-etched nano-patterned $Nd_{0.8}Sr_{0.2}NiO_2$ film.

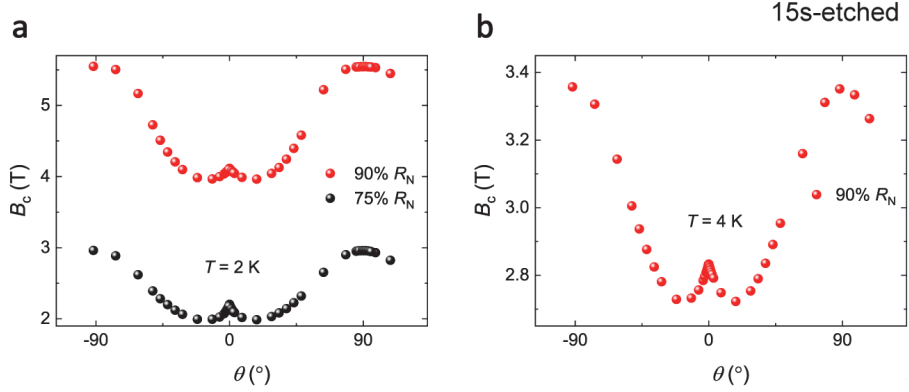

**Fig. S19. $B_c(\theta)$ curves at different temperature in the 15s-etched $Nd_{0.8}Sr_{0.2}NiO_2$ film**. **a, b,** $B_c(\theta)$ curves at $T = 2$ K **(a)**, and 4 K **(b)** using different criterions (i.e., 90% $R_N$ and 75% $R_N$) for determining $B_c$. $B_c(\theta)$ curves show dramatic non-monotonicity and reverse anisotropy with out-of-plane upper critical fields exceeding the in-plane ones.

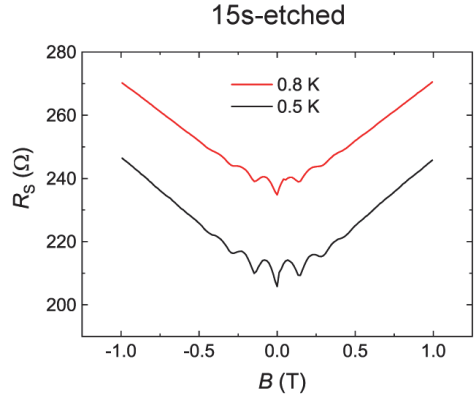

**Fig. S20. Charge-2$e$ quantum oscillations in the 15s-etched $Nd_{0.8}Sr_{0.2}NiO_2$ film**. Charge-2$e$ quantum oscillations in the 15s-etched $Nd_{0.8}Sr_{0.2}NiO_2$ film at different temperatures.

# VII. Additional information for the nano-patterned La$_{0.8}$Sr$_{0.2}$NiO$_2$ film.

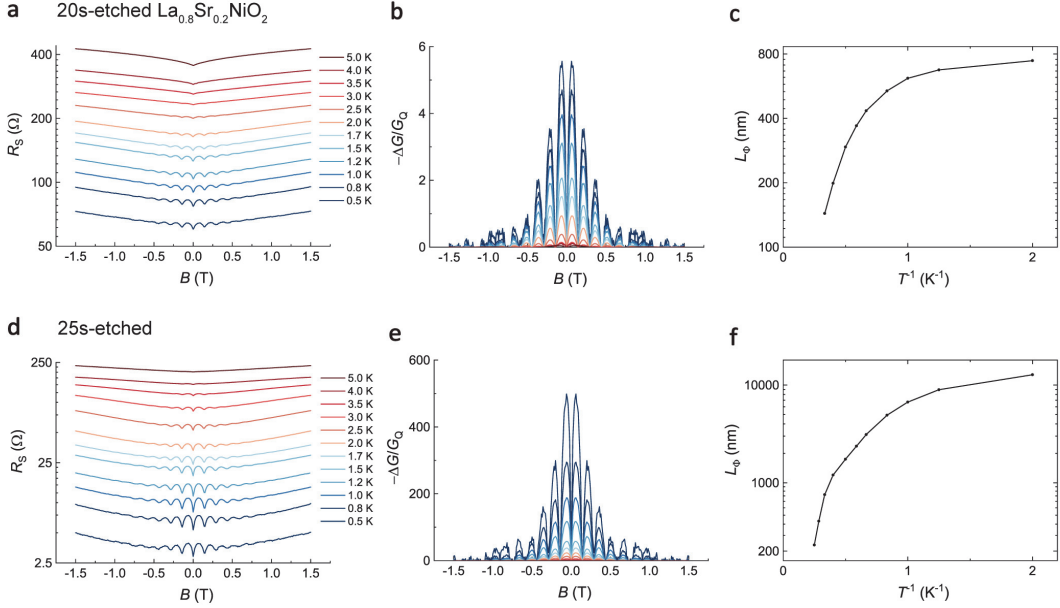

**Fig. S21. Charge-2$e$ quantum oscillations in the nano-patterned La$_{0.8}$Sr$_{0.2}$NiO$_2$ films. a-c,** Quantum oscillations of magnetoresistance (**a**), quantum oscillations of background-subtracted magnetoconductance (**b**, derived from **a**), and temperature-dependent phase coherence length $L_\Phi$ (**c**, derived from **b**) in the 20s-etched nano-patterned La$_{0.8}$Sr$_{0.2}$NiO$_2$ film. **d-f,** Quantum oscillations of magnetoresistance (**d**), quantum oscillations of background-subtracted magnetoconductance (**e**, derived from **d**), and temperature-dependent phase coherence length $L_\Phi$ (**f**, derived from **e**) in the 25s-etched nano-patterned La$_{0.8}$Sr$_{0.2}$NiO$_2$ film.

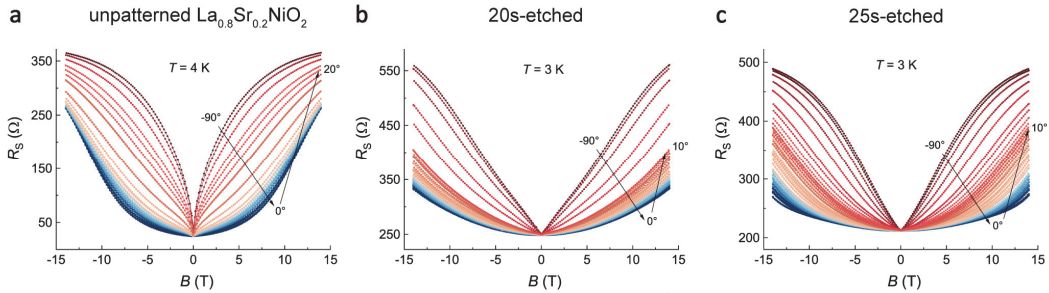

**Fig. S22. $R_S(B)$ curves at different polar angle $\theta$ for the determination of $B_c(\theta)$.** $R_S(B)$ curves with magnetic field applied along different polar angle $\theta$ for unpatterned La$_{0.8}$Sr$_{0.2}$NiO$_2$ film at $T$ = 4 K (**a**), 20s-etched La$_{0.8}$Sr$_{0.2}$NiO$_2$ film at $T$ = 3 K (**b**), and 25s-etched La$_{0.8}$Sr$_{0.2}$NiO$_2$ film at $T$ = 3 K (**c**).